\documentclass[12pt]{iopart}
\eqnobysec
\usepackage{iopams} 
\usepackage{amsthm} 
\usepackage{graphicx}
\usepackage{cite}

\newcommand{\Sw}{{\cal S}(\mathbb R ^+ \! \frac{\ }{\ } \{ a,b \} )}

\begin{document}

\def\llra{\relbar\joinrel\longrightarrow}              
\def\mapright#1{\smash{\mathop{\llra}\limits_{#1}}}    
\def\mapup#1{\smash{\mathop{\llra}\limits^{#1}}}     
\def\mapupdown#1#2{\smash{\mathop{\llra}\limits^{#1}_{#2}}} 

\title[The RHS approach to the Lippmann-Schwinger equation II]{The rigged 
Hilbert space approach to the Lippmann-Schwinger equation. Part II: The 
analytic continuation of the Lippmann-Schwinger bras and kets}

\author{Rafael de la Madrid}
\address{Department of Physics, University of 
California at San Diego, La Jolla, CA 92093 \\
E-mail: {\texttt{rafa@physics.ucsd.edu}}}

\begin{abstract}
The analytic continuation of the Lippmann-Schwinger bras and kets is obtained
and characterized. It is shown that the natural mathematical setting for 
the analytic continuation of the solutions of the Lippmann-Schwinger equation 
is the rigged Hilbert space rather than just the Hilbert space. It is also
argued that this analytic continuation entails the imposition of a time
asymmetric boundary condition upon the group time evolution, resulting into a 
semigroup time evolution. Physically, the semigroup time evolution is simply 
a (retarded or advanced) propagator.
\end{abstract}



\section{Introduction}
\label{sec:introduction}

This paper is devoted to construct and characterize
the analytic continuation of the Lippmann-Schwinger bras and
kets, as well as the analytic continuation of the ``in'' and ``out'' wave 
functions. This paper follows
up on Ref.~\cite{LS1}, where we obtained and characterized the solutions of 
the Lippmann-Schwinger equation associated with the energies of the physical
spectrum. We showed in~\cite{LS1} that such solutions are accommodated by the
rigged Hilbert space rather than by the Hilbert space alone. In this paper,
we shall show that the analytic continuation of the Lippmann-Schwinger bras 
and kets is also accommodated by the rigged Hilbert space rather than by the 
Hilbert space alone.

It was shown in Ref.~\cite{LS1} that the Lippmann-Schwinger bras and
kets are distributions that act on a space of test functions
$\mathbf \Phi \equiv \Sw$. The space $\mathbf \Phi$ arises from invariance
under the action of the Hamiltonian 
and from the need to tame purely imaginary exponentials. These two 
requirements force the functions of $\mathbf \Phi$ to have
a polynomial falloff at infinity. The resulting $\mathbf \Phi$ is a space of 
test functions of the Schwartz type. In this paper, it is shown that the 
analytic continuation of the Lippmann-Schwinger bras and kets are distributions
that act on a space of test functions ${\mathbf \Phi}_{\rm exp}$. The space
${\mathbf \Phi}_{\rm exp}$ arises from invariance under the action of the
Hamiltonian and from the need to tame real exponentials. These two requirements
force the elements of ${\mathbf \Phi}_{\rm exp}$ to fall off at infinity
faster than real exponentials. More precisely, we shall ask the elements of 
${\mathbf \Phi}_{\rm exp}$ to fall off faster than Gaussians. The resulting 
${\mathbf \Phi}_{\rm exp}$ is therefore of the ultra-distribution type. We
recall that an ultra-distribution is an infinitely differentiable test 
function that falls off at infinity faster than exponentials.

In Ref.~\cite{LS1}, we obtained the time evolution of wave functions and of 
the Lippmann-Schwinger bras and kets associated with real energies, and we 
saw that it is given by the standard quantum mechanical group time 
evolution. In this paper, we shall see that analytically continuing the 
time evolution of the wave functions results into a semigroup. We 
shall argue, although not fully prove, that analytically continuing the 
time evolution of Lippmann-Schwinger bras and kets also results 
into a semigroup.

As in Ref.~\cite{LS1}, we restrict ourselves to the spherical shell potential
\begin{equation}
        V({\bf x})\equiv V(r)=\left\{ \begin{array}{ll}
                                0   &0<r<a  \\
                                V_0 &a<r<b  \\
                                0   &b<r<\infty 
                  \end{array} 
                 \right. 
	\label{sbpotential}
\end{equation}
for zero angular momentum. Nevertheless, our results are valid for a larger
class of potentials that include, in particular, potentials of finite 
range. The reason why our results are valid for such a large class of 
potentials
is that, ultimately, such results depend on whether one can analytically
continue the Jost and scattering functions into the whole complex plane. Since
such continuation is possible for potentials that fall off at infinity faster
than any exponential~\cite{NUSSENZVEIG}, our results remain valid for 
a whole lot of interesting potentials.

In this paper, there will be a change in notation with respect to 
Ref.~\cite{LS1}. In 
Ref.~\cite{LS1}, we used the symbol $\psi ^-$ to denote the ``out'' states,
and $\varphi ^{\rm in}$, $\psi ^{\rm out}$ to denote the asymptotically
free ``in'' and ``out'' states. In the present paper, for the 
sake of brevity, we shall use the symbol $\varphi ^-$ to denote the ``out'' 
wave functions $\psi ^-$, and $\varphi$ to denote any asymptotically free wave 
function such as $\varphi ^{\rm in}$ or $\psi ^{\rm out}$.

Throughout the paper, we shall always use radial analytic continuation, 
because the transformation $z\to \sqrt{z}$ converts a radial path of 
integration into another radial path, while it distorts horizontal paths of 
integration.

Since the physical spectrum of the spherical shell potential is 
$[0,\infty )$, one may wonder if performing analytic 
continuations is somehow inconsistent. To qualm any doubts, we recall that
the $S$ matrix, which is defined and unitary on the physical spectrum, is
routinely continued into the complex plane. Much the same way, one can
continue the wave functions and the Lippmann-Schwinger bras and kets into
complex energies.

An important point is what happens with the self-adjointness of the
Hamiltonian on the space of test functions
${\mathbf \Phi}_{\rm exp}$ (and also on $\mathbf \Phi$). These spaces satisfy
\begin{equation}
      {\mathbf \Phi}_{\rm exp} \subset {\mathbf \Phi} \subset 
        {\cal D}(H) \, , 
\end{equation}
where ${\cal D}(H)$ is the domain on which the Hamiltonian is 
self-adjoint~\cite{LS1}. Thus, on $\mathbf \Phi$ and
${\mathbf \Phi}_{\rm exp}$, the Hamiltonian is not a self-adjoint operator 
but just the restriction of a self-adjoint one.

As shall be shown, the analytically continued Lippmann-Schwinger bras and
kets are eigenvectors of the Hamiltonian with complex eigenvalues, and
one may naturally wonder whether such complex eigenvalues are in conflict with
the self-adjointness of the Hamiltonian, which in
principle forbids any complex eigenvalues. To see how self-adjoint
operators can have complex eigenvalues, let us consider the 1D momentum
operator $P=-\rmi \hbar \rmd /\rmd x$. The eigenfunctions of $P$ are
$\rme ^{\rmi px/\hbar}$ with eigenvalue $p$. The eigenvalue $p$ can in
principle be {\it any} complex number, although of course the physical
spectrum of $P$ is the real line and in the completeness
relation there only appear real $p$. Similarly, the eigenvalue equation
for the spherical shell Hamiltonian is valid for any complex number (if 
additional boundary conditions are not imposed). Needless to say, the 
eigenfunctions of the Hamiltonian with complex eigenvalues are not in
the Hilbert space --they are distributions-- and thus there arises no
conflict with the self-adjointness of the Hamiltonian.

Analytic continuations of the Lippmann-Schwinger equation have also been
performed in~\cite{GADELLA84,BG,GADELLA-ORDONEZ,BRTK,CAGADA,GAGO,DIS} by 
assuming that, in the energy representation, the Lippmann-Schwinger bras 
and kets act on two different spaces of Hardy functions. Contrary 
to~\cite{GADELLA84,BG,GADELLA-ORDONEZ,BRTK,CAGADA,GAGO,DIS}, we shall
not make any a priori assumption. Rather, we shall simply obtain the
analytic continuation and study its properties. As it turns out, the
analytically continued Lippmann-Schwinger bras and kets do not act on spaces 
of Hardy functions. Therefore, our results differ drastically from those 
of~\cite{GADELLA84,BG,GADELLA-ORDONEZ,BRTK,CAGADA,GAGO,DIS}.

The rigged Hilbert space we shall use is very similar to, although not the same
as the rigged Hilbert space used by Bollini {\it et al.}~to describe the
resonance (Gamow) states~\cite{BOLLINI1,BOLLINI2}. There are two major 
differences. First, Bollini {\it et al.}~use a space of test functions that
fall off at infinity faster than exponentials, whereas we shall use test
functions that fall off faster than Gaussians. The advantage of using 
Gaussian falloff is that, as will be discussed elsewhere, one can obtain
meaningful resonance expansions. Second, Bollini {\it et al.}~obtain
many results by using the momentum representation and the Fourier transform, 
whereas the present paper deals with the wave-number representation and
the Fourier-like transforms ${\cal F}_{\pm}$ of Sec.~\ref{sec:wnumrepr}. The 
advantage of the wave-number representation is that in such representation, 
the Hamiltonian acts as a multiplication operator, whereas in the momentum 
representation, the Hamiltonian acts as a complicated integral operator. The
simplicity of the wave-number representation will allow us to go beyond
the results of~\cite{BOLLINI1,BOLLINI2}.

The ultimate goal we want to achieve by analytically continuing the solutions 
of the Lippmann-Schwinger equation is to obtain the resonance
states. Although this point will be treated elsewhere, we want to present 
a brief preview of the results. The resonance states are usually obtained
by solving the Schr\"odigner equation subject to purely outgoing boundary
conditions, but they can also be obtained by analytically continuing 
the Lippmann-Schwinger bras and kets into the resonance energies. The results
of this paper will enable us to do just so, and to obtain some novel
properties of the Gamow states. The resulting Gamow states will turn 
out to be different from the so-called ``Gamow vectors'' of~\cite{BG}.

The structure of the paper is as follows. In Sec.~\ref{sec:wnumrepr}, we 
rewrite the results of Ref.~\cite{LS1}
in terms of the wave number, because the analytic continuation is 
more easily done in terms of the wave number than in terms of the energy.

In Sec.~\ref{sec:analicLSe}, we analytically continue the Lippmann-Schwinger
and the ``free'' eigenfunctions. As well, we characterize the analytic and 
the growth properties of such continued eigenfunctions.

In Sec.~\ref{sec:analicLSbk}, we make use of the eigenfunctions
of Sec.~\ref{sec:analicLSe} to analytically continue the
Lippmann-Schwinger and the ``free'' bras and kets.

In Sec.~\ref{sec:rHsanaly}, we construct the rigged Hilbert spaces that
accommodate the analytically continued bras and kets of 
Sec.~\ref{sec:analicLSbk}, and we use these rigged Hilbert spaces to show that
the analytically continued bras and kets are eigenvectors of the Hamiltonian.

In Sec.~\ref{sec:errHsanaly}, we construct and characterize the wave number 
representation of the rigged Hilbert spaces, bras, kets and wave 
functions. In particular, we characterize the analytic and growth
properties of the analytically continued wave functions in the wave
number representation. By means of Gelfand's and Shilov's $M$ and 
$\Omega$ functions~\cite{GELFANDIII}, we shall see how the exponential falloff 
of the elements of ${\mathbf \Phi}_{\rm exp}$ in the position representation 
limits the growth of those elements in the wave number representation. 

In Sec.~\ref{sec:timeevolu}, we construct the time evolution of the 
analytically continued wave functions, bras and kets. By using the
results of Sec.~\ref{sec:errHsanaly}, we shall see that
the analytic continuation of the group time evolution of the wave
functions entails the imposition
of a time asymmetry that converts the group time evolution into a 
semigroup. Such semigroup is just a (retarded or advanced) propagator. We
shall also argue, although not fully prove, that the time evolution of the
analytically continued Lippmann-Schwinger bras and kets is also given by 
semigroups.

In Sec.~\ref{sec:pmvarepos}, we discuss the relation between time asymmetry 
and the $\pm \rmi \varepsilon$ of the Lippmann-Schwinger equation. Finally, 
in Sec.~\ref{sec:conclusions}, we state our conclusions.

All through this paper, $C$ will denote positive constants, not necessarily 
the same at each appearance.

\section{The wave number representation}
\setcounter{equation}{0}
\label{sec:wnumrepr}

The eigenfunctions of the time independent Schr\"odinger equation depend
explicitly not on the energy $E$ but on the wave number $k$~\cite{LS1},
\begin{equation}
       k = \sqrt{\frac{2m}{\hbar ^2}E\,} \, .
      \label{wavenumber}
\end{equation}
In particular, the Lippmann-Schwinger eigenfunctions and the eigenfunction 
expansions depend explicitly on $k$ rather than on $E$. It is therefore
convenient to rewrite their expressions in terms of $k$ before performing 
analytic continuations.

\subsection{The Lippmann-Schwinger eigenfunctions in terms of the (positive) 
wave number}

We start by writing the regular solution in terms of $k$:
\begin{equation}
      \chi (r;k)=  \chi (r;E) = \left\{ \begin{array}{lll}
                   \sin (kr) \quad &0<r<a  \\
                {\cal J}_1(k) \rme ^{\rmi \kappa r}+
                {\cal J}_2(k) \rme ^{-\rmi \kappa r}  \quad  &a<r<b \\
               {\cal J}_3(k) \rme ^{\rmi kr}+{\cal J}_4(k)\rme ^{-\rmi kr}
                                  \quad  &b<r<\infty \, , 
               \end{array} 
                 \right.   
        \label{chiwna}
\end{equation}
where
\begin{equation}
      \kappa =\sqrt{\frac{2m}{\hbar ^2}(E-V_0)\,}=
           \sqrt{k^2-\frac{2m}{\hbar ^2}V_0\,} \ .
     \label{qwavnuem}
\end{equation}
In terms of $k$, the Lippmann-Schwinger eigenfunctions read as
\begin{equation}
     \chi ^{\pm}(r;E)= \sqrt{\frac{1}{\pi} \frac{2m/\hbar ^2}{k}\,} \,  
     \frac{\chi (r;k)}{{\cal J}_{\pm}(k)} \, .
       \label{krepmeigndu}
\end{equation}
The eigenfunctions $\chi ^{\pm}(r;E)$ are $\delta$-normalized as functions
of $E$:
\begin{equation}
      \int_0^{\infty}\rmd r \, \overline{\chi ^{\pm}(r;E)}\chi ^{\pm}(r;E')=
          \delta(E-E') \, .
\end{equation}
The Lippmann-Schwinger eigenfunctions that are $\delta$-normalized 
as functions of $k$ are given by
\begin{equation}
      \chi ^{\pm}(r;k):=\sqrt{\frac{\hbar ^2}{2m}2k\,}\, \chi ^{\pm}(r;E)=
                       \sqrt{\frac{2}{\pi}\,}\, 
                      \frac{\chi (r;k)}{{\cal J}_{\pm}(k)} \, .
      \label{defiphi+-}
\end{equation}
Indeed, it is easy to check that
\begin{equation}
      \int_0^{\infty}\rmd r \, \overline{\chi ^{\pm}(r;k)}\chi ^{\pm}(r;k')=
          \delta(k-k') \, .
\end{equation}

\subsection{The ``in'' and ``out'' bras, kets and wave functions in terms of 
the (positive) wave number}

Once we have expressed the Lippmann-Schwinger eigenfunctions as 
$\delta$-normalized eigenfunctions of $k$, we can construct the unitary 
operators that transform from the position into the wave number 
representation. These operators will be denoted by ${\cal F}_{\pm}$. We shall
also rewrite the Lippmann-Schwinger bras and kets, along with the basis
expansions induced by them, in terms of $k$.

We first define the wave number representation, $\widehat{f}(k)$, of any 
function $\widehat{f}(E)$ in $L^2([0,\infty ),\rmd E)$ by
\begin{equation}
      \widehat{f}(k):=\sqrt{\frac{\hbar ^2}{2m}\, 2k\,}\, \widehat{f}(E) \, .
      \label{wnresifunc}
\end{equation}
Because $\widehat{f}(E)$ belongs to $L^2([0,\infty ),\rmd E)$, the function 
$\widehat{f}(k)$ belongs to $L^2([0,\infty ),\rmd k)$. The expressions for 
${\cal F}_{\pm}$ and ${\cal F}_{\pm}^{-1}$ as integral operators can be 
easily obtained from the expressions for the operators 
$U_{\pm}$ and $U_{\pm}^{-1}$ of Ref.~\cite{LS1} with help from 
Eqs.~(\ref{wavenumber}), (\ref{defiphi+-}) and (\ref{wnresifunc}):
\numparts
\begin{equation}
       \widehat{f}_{\pm}(k)=({\cal F}_{\pm}f)(k)=\int_0^{\infty}\rmd r \,
        f(r)\overline{\chi ^{\pm}(r;k)} \, ,
       \label{Fpm}
\end{equation}
\begin{equation}
      f(r)= ({\cal F}_{\pm}^{-1}\widehat{f}_{\pm})(r)=\int_0^{\infty}\rmd k \, 
                       \widehat{f}_{\pm}(k)\chi ^{\pm}(r;k) \, .
      \label{iFpm}
\end{equation}
\endnumparts
By construction, ${\cal F}_{\pm}$ are unitary operators from 
$L^2([0,\infty ),\rmd r)$ onto $L^2([0,\infty ),\rmd k)$:
\begin{equation}
     \begin{array}{rcl}
      {\cal F}_{\pm}:L^2([0,\infty ),\rmd r) 
      &\longmapsto & L^2( [0,\infty ),\rmd k)  \\
       f(r)& \longmapsto & 
         \widehat{f}_{\pm}(k)=({\cal F}_{\pm}f )(k) \, .
    \end{array} 
      \label{F+-slthe}
\end{equation}
The notation ${\cal F}_{\pm}$ intends to stress that ${\cal F}_{\pm}$ are 
Fourier-like transforms.

In terms of $k$, the Lippmann-Schwinger bras and kets become
\numparts
\begin{equation}
      \langle ^{\pm}k| = \sqrt{\frac{\hbar ^2}{2m}\, 2k\,} \, \langle ^{\pm}E|
          \, , \quad k>0 \, ,
\end{equation}
\begin{equation}
      |k^{\pm}\rangle = \sqrt{\frac{\hbar ^2}{2m}\, 2k\,} \, |E^{\pm}\rangle 
          \, , \quad k>0 \, ;
\end{equation}
\endnumparts
that is,
\numparts
\begin{equation}
       \langle ^{\pm}k| \varphi ^{\pm}\rangle =
         \int_0^{\infty} \rmd r \,  
    \langle ^{\pm}k| r\rangle \langle r | \varphi ^{\pm}\rangle  \, , 
                 \quad k>0 \, ,
        \  \varphi ^{\pm} \in {\mathbf \Phi} \, ,
         \label{brak+-}  
\end{equation}
\begin{equation}
       \langle \varphi ^{\pm}|k^{\pm}\rangle =
         \int_0^{\infty} \rmd r \,  
           \langle \varphi ^{\pm}|r\rangle \langle r|k^{\pm}\rangle \, , 
             \quad k>0 \, ,
        \  \varphi ^{\pm} \in {\mathbf \Phi} \, ,
        \label{ketk+-} 
\end{equation}
\endnumparts
where ${\mathbf \Phi} \equiv \Sw$ is the Schwartz-like space built 
in~\cite{LS1} and
\numparts
\begin{equation}
      \langle r|k^{\pm}\rangle = \chi ^{\pm}(r;k)  \, , \quad k>0 \, , 
\end{equation}
\begin{equation}
      \langle ^{\pm}k|r \rangle = \overline{\chi ^{\pm}(r;k)} =     
            \chi ^{\mp}(r;k)      \, , \quad k>0 \, . 
     \label{leftLSe+-}
\end{equation}
\endnumparts

Using the corresponding formal identity for the bras and kets in terms of
$E$, one can express the identity operator as
\begin{equation}
      1= \int_0^{\infty}\rmd k \, |k^{\pm} \rangle \langle ^{\pm}k| \, ; 
           \label{residenpm}
\end{equation}
that is,
\begin{equation}
       \langle r| \varphi ^{\pm}\rangle =
         \int_0^{\infty} \rmd k \,  
    \langle r|k^{\pm}\rangle \langle ^{\pm}k | \varphi ^{\pm}\rangle  \, , 
                   \quad k>0 \, ,
        \  \varphi ^{\pm} \in {\mathbf \Phi} \, . 
\end{equation}
One can also express the $S$-matrix element as
\begin{equation}
      (\varphi ^-, \varphi ^+) = \int_0^{\infty}\rmd k \,  
     \langle \varphi ^-|k^- \rangle S(k) \langle ^+k|\varphi ^+\rangle \, , 
         \quad  \varphi ^{\pm} \in {\mathbf \Phi} \, ,
\end{equation}
where 
\begin{equation}
       S(k)=\frac{{\cal J}_-(k)}{{\cal J}_+(k)}  \, .
\end{equation}

Since in the energy representation $H$ acts as multiplication by $E$,
in the wave number representation $H$ acts as multiplication by 
$\frac{\hbar ^2}{2m}k^2$:
\begin{equation}
      (\widehat{H} \widehat{f})(k)= 
        ({\cal F}_{\pm}H{\cal F}_{\pm}^{\dagger}\widehat{f})(k) =
         \frac{\hbar ^2}{2m}k^2 \,  \widehat{f}(k) \, .
       \label{wnrofHamil}
\end{equation}
As well, the bras $\langle ^{\pm}k|$ and kets $|k^{\pm}\rangle$ are, 
respectively, left and right eigenvectors of $H$ with eigenvalue 
$\frac{\hbar ^2}{2m}k^2$:
\begin{equation}
       \langle ^{\pm}k| H =\frac{\hbar ^2}{2m}k^2  \langle ^{\pm}k| \, ,
\end{equation}
\begin{equation}
       H |k^{\pm}\rangle =\frac{\hbar ^2}{2m}k^2  |k^{\pm}\rangle \, .
\end{equation}

\subsection{The ``free'' bras, kets and wave functions in terms of the 
(positive) wave number}

The expressions for the eigenfunctions, bras and kets of the free 
Hamiltonian $H_0$ can also be rewritten in terms of $k$.

The ``free'' eigenfunction that is $\delta$-normalized as a function of $k$ 
is given by
\begin{equation}
      \chi _0(r;k):=\sqrt{\frac{\hbar ^2}{2m}2k\,}\, \chi _0(r;E)=
                       \sqrt{\frac{2}{\pi}\,}\, \sin (kr) \, .
      \label{defiphi0}
\end{equation}
By using Eqs.~(\ref{wavenumber}), (\ref{wnresifunc}) and (\ref{defiphi0}),
together with the expression for the integral operator $U_0$ obtained
in Ref.~\cite{IJTP03}, one can construct the following integral operator and 
its inverse:
\numparts
\begin{equation}
       \widehat{f}_{0}(k)=({\cal F}_{0}f)(k)=\int_0^{\infty}\rmd r \,
        f(r)\overline{\chi _0(r;k)} \, ,
       \label{F0}
\end{equation}
\begin{equation}
      f(r)= ({\cal F}_{0}^{-1}\widehat{f}_{0})(r)=\int_0^{\infty}\rmd k \, 
                       \widehat{f}_{0}(k)\chi _0(r;k) \, .
      \label{iF0}
\end{equation}
\endnumparts
The transform ${\cal F}_{0}$ is a unitary operator from 
$L^2([0,\infty ),\rmd r)$ onto $L^2([0,\infty ),\rmd k)$:
\begin{equation}
     \begin{array}{rcl}
      {\cal F}_{0}:L^2([0,\infty ),\rmd r) 
      &\longmapsto & L^2( [0,\infty ),\rmd k)  \\
       f(r)& \longmapsto & 
         \widehat{f}_{0}(k)=({\cal F}_{0}f )(k) \, .
    \end{array} 
      \label{F0slthe}
\end{equation}

In terms of $k$, the ``free'' bras and kets become
\numparts
\begin{equation}
      \langle k| = \sqrt{\frac{\hbar ^2}{2m}\, 2k\,} \, \langle E|
          \, , \quad k>0 \, ,
\end{equation}
\begin{equation}
      |k\rangle = \sqrt{\frac{\hbar ^2}{2m}\, 2k\,} \, |E\rangle 
          \, , \quad k>0 \, ;
\end{equation}
\endnumparts
that is,
\numparts
\begin{equation}
       \langle k| \varphi \rangle =
         \int_0^{\infty} \rmd r \,  \langle k| r\rangle 
           \langle r | \varphi \rangle  \, , \quad k>0 \, ,
        \label{bra0} 
\end{equation}
\begin{equation}
       \langle \varphi |k\rangle =
         \int_0^{\infty} \rmd r \,  
           \langle \varphi |r\rangle \langle r|k\rangle \, , 
                 \quad k>0  \, , 
        \label{ket0}
\end{equation}
\endnumparts
where
\numparts
\begin{equation}
      \langle k|r \rangle = \overline{\chi _0(r;k)} = \chi _0(r;k)          
                     \, , \quad k>0 \, , 
      \label{lefteigen0}
\end{equation}
\begin{equation}
      \langle r|k\rangle = \chi _0(r;k)  \, , \quad k>0 \, ,
        \label{righteigen0} 
\end{equation}
\endnumparts
and where $\varphi$ denotes either $\varphi ^{\rm in}$ or $\psi ^{\rm out}$.

Using the corresponding formal identity for the ``free'' bras and kets in 
terms of $E$, one can express the identity operator as
\begin{equation}
      1 = \int_0^{\infty}\rmd k \, |k \rangle \langle k| \, ; 
\end{equation}
that is,
\begin{equation}
       \langle r| \varphi \rangle =
         \int_0^{\infty} \rmd k \,  
    \langle r|k\rangle \langle k | \varphi \rangle  \, , 
          \quad k>0 \, .
        \label{0resolind}        
\end{equation}

In the wave number representation $H_0$ acts as multiplication by 
$\frac{\hbar ^2}{2m}k^2$:
\begin{equation}
      (\widehat{H}_0 \widehat{f})(k)= 
        ({\cal F}_{0}H_0{\cal F}_{0}^{\dagger}\widehat{f}_0)(k) =
         \frac{\hbar ^2}{2m}k^2 \,  \widehat{f}_0(k) \, .
\end{equation}
As well, the bras $\langle k|$ and kets $|k\rangle$ are, 
respectively, left and right eigenvectors of $H_0$ with eigenvalue 
$\frac{\hbar ^2}{2m}k^2$:
\begin{equation}
       \langle k| H_0 =\frac{\hbar ^2}{2m}k^2  \langle k| \, ,
\end{equation}
\begin{equation}
       H_0 |k\rangle =\frac{\hbar ^2}{2m}k^2  |k\rangle \, .
\end{equation}

Finally, the M{\o}ller operators $\Omega _{\pm}$ can be expressed in terms of 
the operators ${\cal F}_{\pm}$ and ${\cal F}_0$ as
\begin{equation}
      \Omega _{\pm}={\cal F}_{\pm}^{\dagger}{\cal F}_0 \, , 
\end{equation}
and they connect the ``free'' with the ``in'' and ``out'' kets by
\begin{equation}
      \Omega _{\pm}|k\rangle =|k^{\pm}\rangle \, , \quad k>0 \, .
         \label{omegak}
\end{equation}

\section{The analytic continuation of the Lippmann-Schwinger eigenfunctions}
\setcounter{equation}{0}
\label{sec:analicLSe}

Equations~(\ref{chiwna})-(\ref{omegak}), in particular the 
expressions for the Lippmann-Schwinger eigenfunctions, were obtained in 
Ref.~\cite{LS1} by means of the Sturm-Liouville theory and are valid when 
$E$ and $k$ are positive.\footnote{It is somewhat remarkable
that the Sturm-Liouville theory actually uses complex energies, although it 
makes do with a particular branch of the square root function instead of 
a Riemann surface.} We are now going to perform 
the (radial) analytic continuation of the Lippmann-Schwinger eigenfunctions 
into the complex plane. Equation~(\ref{wavenumber}) provides the Riemann 
surface for such analytic continuation. 

\begin{figure}[ht]
\hskip-0.5cm \includegraphics{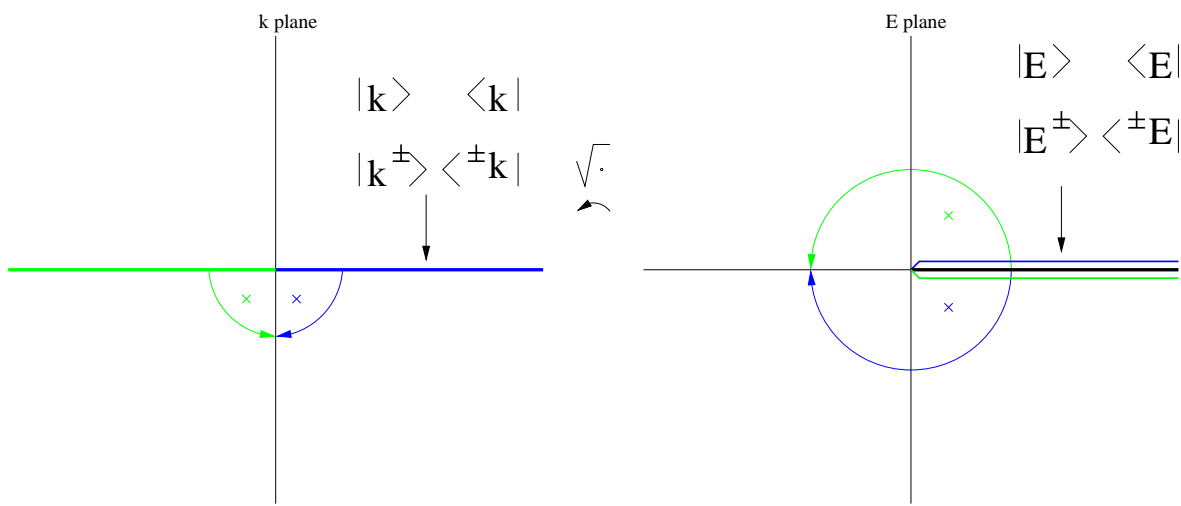}
\caption{The boundary values of the Lippmann-Schwinger and of the ``free'' 
bras and kets.}
\label{fig:lsur}
\end{figure}

The analytic continuation of $\chi ^{\pm}(r;E)$ is obtained in two 
steps. First, one specifies the boundary values of the 
Lippmann-Schwinger eigenfunctions on the upper rim of the cut. And second, 
one continues those boundary values into the whole two-sheeted Riemann
surface, see Fig.~\ref{fig:lsur}. The boundary values of the 
Lippmann-Schwinger eigenfunctions on the upper rim are given by 
Eq.~(\ref{krepmeigndu}).

Because the $\chi ^{\pm}(r;E)$ depend explicitly on $k$ rather than on $E$, 
the analytic continuation of the Lippmann-Schwinger eigenfunctions is more 
easily obtained in terms of $k$, i.e., in terms of the eigenfunctions
$\chi ^{\pm}(r;k)$. The $E$-continuation described above
translates into a $k$-continuation as follows. First, one specifies
the boundary values that the Lippmann-Schwinger eigenfunctions take on the 
positive $k$-axis. And second, one continues those boundary values into 
the whole $k$-plane. Since the boundary values of the Lippmann-Schwinger 
eigenfunctions on the positive $k$-axis are given by Eq.~(\ref{defiphi+-}),
and since $\chi ^{\pm}(r;k)$ are expressed in terms of well-known analytic 
functions, the continuation of $\chi ^{\pm}(r;k)$ from the positive $k$-axis 
into the whole wave-number plane is well defined.

Obviously, the analytic continuation of the ``free'' eigenfunctions 
$\chi _0 (r;k)$ follows the same procedure.

A word on notation. Whenever they become complex, we shall
denote the energy $E$ and the wave number $k$ by respectively $z$ and 
$q$. Accordingly, the
continuations of $\chi ^{\pm}(r;E)$, $\chi _0(r;E)$ and $\chi ^{\pm}(r;k)$,
$\chi _0(r;k)$ will be denoted by $\chi ^{\pm}(r;z)$, $\chi _0(r;z)$ and 
$\chi ^{\pm}(r;q)$, $\chi _0(r;q)$. In bra-ket notation, the analytically
continued eigenfunctions will be written as
\begin{eqnarray}
       \langle r|q^{\pm}\rangle =\chi ^{\pm}(r;q) \, ,  \\
       \langle ^{\pm}q|r \rangle =\chi ^{\mp}(r;q) \, ,  \\
       \langle r|q\rangle =\chi _0(r;q) \, ,  \\
       \langle q|r \rangle =\chi _0(r;q) \, .
\end{eqnarray}
In appendix~\ref{sec:apusfour}, we list several useful relations satisfied 
by these analytically continued eigenfunctions.

In doing analytic continuations, it is important to keep in mind that the
combined operations of analytic continuation and complex conjugation do
not commute (and also differ in whether the resulting function is analytic
or not). The reason lies in the fact that if $f(z)$ is an analytic function,
then $\overline{f(z)}$ is not an analytic function. This is why the analytic 
continuation of $\overline{f(E)}$ must in general be written as 
$\overline{f(\overline{z})}$. For example, for real wave numbers it holds that
\begin{equation}
      \chi ^+(r;k)= \overline{\chi ^-(r;k)} \, .
           \label{exampans}
\end{equation}
When we analytically continue Eq.~(\ref{exampans}), we must write
\begin{equation}
      \chi ^+(r;q)= \overline{\chi ^-(r;\overline{q})} \, ,
\end{equation}
rather than
\begin{equation}
      \chi ^+(r;q)= \overline{\chi ^-(r;q)} \, ,
        \label{falseanald}
\end{equation}
since $\overline{\chi ^-(r;q)}$ is not analytic. What is more,
Eq.~(\ref{falseanald}) is false.

We now turn to characterize the analytic and the growth 
properties of $\chi ^{\pm}(r;q)$. Such properties will be needed in the 
next section. In order to characterize the analytic properties of 
$\chi ^{\pm}(r;q)$, we define the following sets:
\begin{equation}
     Z_{\pm} = \{ q \in {\mathbb C} \, | \ {\cal J}_{\pm}(q) = 0 \} \, .
\end{equation}
The set $Z_{\pm}$ contains the zeros of the Jost function 
${\cal J}_{\pm}(q)$. Because of Eq.~(\ref{jsejmqmps}), a wave number
$q$ belongs to $Z_+$ if, and only if, $-q$ belongs to $Z_-$. The elements of
$Z_+$ are simply the discrete, denumerable poles of the $S$ matrix. Since 
$\chi (r;q)$ and ${\cal J}_{\pm}(q)$ are analytic in the 
whole $k$-plane~\cite{NUSSENZVEIG,TAYLOR}, 
$\chi ^{\pm}(r;q)$ is analytic in the whole $k$-plane except at
$Z_{\pm}$, where its poles are located.

In order to characterize the growth of $\chi ^{\pm}(r;q)$, we study
first the growth of $\chi (r;q)$. The growth of 
$\chi (r;q)$ is bounded by the following estimate 
(see, for example, Eq.~(12.6) in Ref.~\cite{TAYLOR}):
\begin{equation}
      \left| \chi (r;q)\right| \leq C \, 
        \frac{\left|q\right|r}{1+\left|q\right|r} \,  
      \rme ^{|{\rm Im}(q)|r} \, , \quad q\in {\mathbb C} \, .   
      \label{boundrs}
\end{equation}
From Eqs.~(\ref{defiphi+-}) and 
(\ref{boundrs}), it follows that the eigenfunctions $\chi ^{\pm}(r;q)$ 
satisfy
\begin{equation}
    \hskip-1cm  \left| \chi ^{\pm}(r;q) \right| \leq 
         \frac{C}{|{\cal J}_{\pm}(q)|} \, 
      \frac{|q|r}{1+|q|r} \,
      \rme ^{|{\rm Im}(q)|r }  \, . 
     \label{estimateofphi}
\end{equation}
When $q \in Z_{\pm}$, the Lippmann-Schwinger eigenfunction $\chi ^{\pm}(r;q)$ 
blows up to infinity.

We can further refine the estimates~(\ref{estimateofphi}) by
characterizing the growth of $1/|{\cal J}_{\pm}(q)|$ in different regions
of the complex plane. The following proposition, which is based on
well-known results~\cite{TAYLOR,NUSSENZVEIG}, and whose proof can be found in 
appendix~\ref{sec:A3}, characterizes the growth
of $1/|{\cal J}_{\pm}(q)|$ in different regions of the $k$-plane for the 
spherical shell potential:

\vskip0.5cm

\newtheorem*{Prop1}{Proposition~1}
\begin{Prop1} \label{Prop1} The inverse of the Jost function 
${\cal J}_{+}(q)$ is bounded in the upper half of the complex wave-number 
plane:
\begin{equation}
    \frac{1}{\left| {\cal J}_{+}(q) \right|} \leq C \, ,
    \quad    {\rm Im}(q)\geq 0   \, .
    \label{boundinjslp+}
\end{equation} 
In the lower half-plane, $\frac{1}{{\cal J}_{+}(q)}$ is infinite whenever 
$q \in Z_+$. As $|q|$ tends to $\infty$ in the lower half plane, 
we have
\begin{equation}
    \frac{1}{{\cal J}_{+}(q)} \approx 
    \frac{1}{1 - C q^{-2} \rme ^{2\rmi q b}} \equiv 
       \frac{1}{\lambda (q)} \, , \quad
       (|q|\to \infty \, , \  {\rm Im}(q)<0)   \, .
    \label{boundinjsup+}
\end{equation} 
The above estimates are satisfied by ${\cal J}_{-}(q)$ when we exchange the 
upper for the lower half plane, and $Z_+$ for $Z_-$:
\begin{equation}
    \frac{1}{\left| {\cal J}_{-}(q) \right|} \leq C \, ,
    \quad {\rm Im}(q)\leq 0   \, .
    \label{boundinjslp-}
\end{equation} 
\begin{equation}
    \frac{1}{{\cal J}_{-}(q)} \approx
     \frac{1}{1 - C q^{-2} \rme ^{-2\rmi q b}} \equiv
      \frac{1}{\lambda (-q)}
            \, , \quad
            (|q|\to \infty \, , \  {\rm Im}(q)>0) \, .
    \label{boundinjsup-}
\end{equation} 
\end{Prop1}

\vskip0.5cm

Equation~(\ref{estimateofphi}) and Proposition~1 imply, in particular, that
the growth of the ``out'' eigenfunction in the lower half plane is limited
by
\begin{equation}
      \left| \chi ^-(r;q)\right| \leq C \, 
        \frac{\left|q\right|r}{1+\left|q\right|r} \,  
      \rme ^{|{\rm Im}(q)|r} \, , \quad  {\rm Im}(q)\leq 0  \, .   
      \label{boundrschi-lower}
\end{equation}

To finish this section, we recall that the ``free'' eigenfunctions are 
analytic in the whole complex plane and 
satisfy an estimate similar to that in Eq.~(\ref{boundrs}), as shown by 
Eq.~(12.4) in Ref.~\cite{TAYLOR}:
\begin{equation}
        |\chi _0(r;q)| = | \sqrt{2/ \pi \,} \sin (qr) | 
           \leq C \frac{|q|r}{1+|q|r} \, \rme ^{|{\rm Im}(q)|r} 
          \, , \quad q\in {\mathbb C}  \, .
    \label{estchiozero}
\end{equation}

\section{The analytic continuation of the Lippmann-Schwinger bras and kets}
\setcounter{equation}{0}
\label{sec:analicLSbk}

The analytic continuation of the Lippmann-Schwinger bras~(\ref{brak+-})
is defined for any complex wave number $q$ in the
distributional way:
\begin{equation}
      \begin{array}{rcl} 
    \langle ^{\pm}q| :{\mathbf \Phi}_{\rm exp} &  \longmapsto & {\mathbb C} \\
       \varphi  ^{\pm} & \longmapsto & \langle ^{\pm}q|\varphi ^{\pm}\rangle =
       \int_0^{\infty}\rmd r\, \varphi ^{\pm} (r) \chi ^{\mp}(r;q) \, ,
       \end{array}
       \label{LSdefinitionbra+-q}
\end{equation}
where the functions $\varphi ^{\pm}(r)$ belong to a space of test functions
${\mathbf \Phi}_{\rm exp}$ that will be constructed in the next 
section. In the bra-ket notation, Eq.~(\ref{LSdefinitionbra+-q}) can be recast
as
\begin{equation}
     \langle ^{\pm}q|\varphi ^{\pm}\rangle =
       \int_0^{\infty}\rmd r\,  \langle ^{\pm}q|r\rangle 
        \langle r|\varphi ^{\pm}\rangle \, .
       \label{LSdefinitionbra+-qDn}
\end{equation}
Obviously, when the complex wave number $q$ tends to the real, positive
wave number $k$, the bras $\langle ^{\pm}q|$ tend to the bras 
$\langle ^{\pm}k|$. 

Similarly to the bras~(\ref{brak+-}), the analytic continuation of the 
Lippmann-Schwinger kets~(\ref{ketk+-}) is defined as
\begin{equation}
       \begin{array}{rcl}
  |q^{\pm}\rangle :{\mathbf \Phi}_{\rm exp} &  \longmapsto & {\mathbb C}  \\
       \varphi  ^{\pm} & \longmapsto & \langle \varphi ^{\pm}|q^{\pm}\rangle =
       \int_0^{\infty}\rmd r\, 
              \overline{\varphi ^{\pm} (r)} \chi ^{\pm}(r;q) \, ,
      \end{array}
       \label{LSdefinitionket+-q}
\end{equation}
which in bra-ket notation becomes
\begin{equation}
    \langle \varphi ^{\pm}|q^{\pm}\rangle =
       \int_0^{\infty}\rmd r\, \langle \varphi ^{\pm}|r\rangle 
        \langle r|q^{\pm}\rangle  \, .
           \label{LSdefinitionket+-qDn}
\end{equation}
By construction, when $q$ tends to $k$, the kets $|q^{\pm}\rangle$ tend to 
the kets $|k^{\pm}\rangle$. 

The bras~(\ref{LSdefinitionbra+-q}) and kets~(\ref{LSdefinitionket+-q}) are 
defined for all complex $q$ except at those $q$ at which the 
corresponding eigenfunction has a pole. Hence, $\langle ^{-}q|$ and 
$|q^{+}\rangle$ are defined everywhere except in $Z_+$,
whereas $\langle ^{+}q|$ and $|q^{-}\rangle$ are defined everywhere except 
in $Z_-$. At those poles, one can still define bras and kets if in 
definitions~(\ref{LSdefinitionbra+-q}) and 
(\ref{LSdefinitionket+-q}) one substitutes the eigenfunctions 
$\chi ^{\pm}(r;q)$ by their residues at the pole:
\begin{equation}
      \begin{array}{rcl} 
    \langle ^{\pm}q| :{\mathbf \Phi}_{\rm exp} &  \longmapsto & {\mathbb C} \\
       \varphi  ^{\pm} & \longmapsto & \langle ^{\pm}q|\varphi ^{\pm}\rangle =
       \int_0^{\infty}\rmd r\, \varphi ^{\pm} (r) \, 
                  {\rm res}[\chi ^{\mp}(r;q)] \, , \quad q\in Z_{\mp} \, ,
       \end{array}
       \label{LSdefinitionbra+-qres}
\end{equation}
\begin{equation}
       \begin{array}{rcl}
  |q^{\pm}\rangle :{\mathbf \Phi}_{\rm exp} &  \longmapsto & {\mathbb C}  \\
       \varphi  ^{\pm} & \longmapsto & \langle \varphi ^{\pm}|q^{\pm}\rangle =
       \int_0^{\infty}\rmd r\, 
              \overline{\varphi ^{\pm} (r)} \, {\rm res}[\chi ^{\pm}(r;q)] \, ,
          \quad q\in Z_{\pm} \, .
      \end{array}
       \label{LSdefinitionket+-qres}
\end{equation}
In this way, one can 
associate bras $\langle ^{\pm}q|$ and kets $|q^{\pm}\rangle$ with every 
complex wave number $q$.

The analytic continuation of the ``free'' bras and kets~(\ref{bra0}) and 
(\ref{ket0}) into any complex wave number $q$ is defined in the obvious way:
\begin{equation}
     \langle q|\varphi \rangle =
       \int_0^{\infty}\rmd r\,  \langle q|r\rangle 
        \langle r|\varphi \rangle
         \int_0^{\infty}\rmd r\, \varphi (r) \chi _0(r;q) 
              \, , \qquad  q \in {\mathbb C} \, ,
         \label{anconfbra} 
\end{equation}
\begin{equation}
    \langle \varphi |q\rangle =
       \int_0^{\infty}\rmd r\, \langle \varphi |r\rangle 
        \langle r|q\rangle =
     \int_0^{\infty}\rmd r\, \overline{\varphi (r)} \chi _0(r;q)
           \, , \quad q \in {\mathbb C} \, ,
           \label{anconfket}
\end{equation}
where $\varphi$ denotes any asymptotically free wave function. Likewise 
definitions~(\ref{LSdefinitionbra+-q}) and (\ref{LSdefinitionket+-q}),
definitions~(\ref{anconfbra}) and (\ref{anconfket}) make sense 
when $\varphi$ belongs to ${\mathbf \Phi}_{\rm exp}$. 

From the analytic continuation of the bras and kets into any
complex wave number, one can now obtain the analytic continuation of the 
bras and kets into any complex energy of the Riemann surface:
\begin{equation}
      \begin{array}{ccc}
    |z^{\pm}\rangle =\sqrt{\frac{2m}{\hbar ^2}\frac{1}{2q}}\, |q^{\pm}\rangle
    \, ,  & \quad &  
    \langle ^{\pm}z|=\sqrt{\frac{2m}{\hbar ^2}\frac{1}{2q}}\, \langle ^{\pm}q|
    \, , \\ [3ex]
  |z\rangle =\sqrt{\frac{2m}{\hbar ^2}\frac{1}{2q}} \, |q\rangle 
        \, , & \quad & 
      \langle z|= \sqrt{\frac{2m}{\hbar ^2}\frac{1}{2q}} \, \langle q| \, .
       \end{array}
\end{equation}

\section{Construction of the rigged Hilbert space for the analytic
continuation of the Lippmann-Schwinger bras and kets}
\setcounter{equation}{0}
\label{sec:rHsanaly}

Likewise the bras and kets associated with real energies, the analytic
continuation of the Lippmann-Schwinger bras and kets must be described 
within the rigged Hilbert space rather than just within the Hilbert 
space. We shall denote the rigged Hilbert space for the analytically 
continued bras by
\begin{equation}
     {\mathbf \Phi}_{\rm exp} \subset L^2([0,\infty ), \rmd r) 
      \subset {\mathbf \Phi}_{\rm exp}^{\prime} \, ,
         \label{rhsexpp}
\end{equation} 
and the one for the analytically continued kets by
\begin{equation}
     {\mathbf \Phi}_{\rm exp} \subset L^2([0,\infty ), \rmd r) 
      \subset {\mathbf \Phi}_{\rm exp}^{\times} \, .
     \label{rhsexpt}
\end{equation} 

In principle, we should construct the space of test functions separately for 
the ``in'' and for the ``out'' wave functions. But since they turn out to be 
the same, we present the construction for both cases at once.

The functions $\varphi ^{\pm}\in {\mathbf \Phi}_{\rm exp}$ must satisfy 
the following conditions: 
\numparts
\begin{eqnarray}
  \hskip-1cm && \hskip-0.4cm \begin{array}{lc}
        & \bullet \  \mbox{They belong to the maximal invariant 
                 subspace $\cal D$ of} \ H, \\ [1ex]
       & {\cal D} = \bigcap_{n=0}^{\infty} {\cal D}(H^n) \, .
        \end{array}
           \label{condition1} \\ [2ex]
  \hskip-1cm &&\bullet \ \mbox{They are such that
   definitions~(\ref{LSdefinitionbra+-q}) and (\ref{LSdefinitionket+-q}) 
make sense.} 
         \label{condition2}
\end{eqnarray}
\endnumparts
The reason why $\varphi ^{\pm}$ must satisfy condition~(\ref{condition1})
is that such condition guarantees that all the powers of the Hamiltonian are 
well defined. Condition~(\ref{condition1}), however, is not sufficient to 
obtain well-defined bras and kets associated with complex wave numbers. In 
order for $\langle ^{\pm}q|$ and $|q^{\pm}\rangle$
to be well defined, the wave functions $\varphi ^{\pm}(r)$ must be well 
behaved so the integrals in Eqs.~(\ref{LSdefinitionbra+-q}) and 
(\ref{LSdefinitionket+-q}) converge. How well $\varphi ^{\pm}(r)$ must 
behave is determined by how bad $\chi ^{\pm}(r;q)$ behave. Since by 
Eq.~(\ref{estimateofphi})
$\chi ^{\pm}(r;q)$ grow exponentially with $r$, the wave functions 
$\varphi ^{\pm}(r)$ have to, essentially, tame real exponentials. If we define
\begin{equation}
   \hskip-2cm   \| \varphi ^{\pm}\|_{n,n'} := \sqrt{\int_{0}^{\infty}\rmd r \, 
    \left| \frac{nr}{1+nr}\, \rme ^{nr^2/2} (1+H)^{n'}
             \varphi ^{\pm}(r) \right|^2 \, } 
                 \, , \quad n,n'=0,1,2, \ldots \, , 
      \label{normsLS}
\end{equation}
then the space ${\mathbf \Phi}_{\rm exp}$ is given by
\begin{equation}
      {\mathbf \Phi}_{\rm exp}= 
      \left\{ \varphi ^{\pm}\in {\cal D} \, | 
      \ \| \varphi ^{\pm} \|_{n,n'}<\infty \, , \ n,n'=0,1,2,\ldots \right\} .
       \label{phiexp}
\end{equation}
This is just the space of square integrable functions which belong to the
maximal invariant subspace of $H$ and for which the quantities~(\ref{normsLS})
are finite. In particular, because $\varphi ^{\pm}(r)$ satisfy the 
estimates~(\ref{normsLS}), $\varphi ^{\pm}(r)$ fall off at infinity faster 
than $\rme ^{-r^2}$, that is, their tails fall off faster than Gaussians. 

From Eq.~(\ref{estimateofphi}), it is clear that the integrals in
Eqs.~(\ref{LSdefinitionbra+-q}) and (\ref{LSdefinitionket+-q})
converge already for functions that fall off at infinity faster than any 
exponential. We have imposed Gaussian falloff because it allows us to perform
expansions in terms of the Gamow states, as will
be discussed elsewhere.

It is illuminating to compare the space of test functions needed to
accommodate the Lippmann-Schwinger bras and kets associated with real wave
numbers, the space $\mathbf \Phi$ of Ref.~\cite{LS1}, with the space of test
functions needed to accommodate their analytic continuation, the space 
${\mathbf \Phi}_{\rm exp}$
of Eq.~(\ref{phiexp}). Because for real wave numbers the Lippmann-Schwinger 
eigenfunctions behave like purely imaginary exponentials, in this case we 
only need to impose on the test functions a polynomial falloff, thereby 
obtaining a space of test functions very similar to the Schwartz space. By 
contrast, for complex wave numbers the Lippmann-Schwinger eigenfunctions blow 
up exponentially, and therefore we need to impose on the test functions an 
exponential falloff that damps such an exponential blowup.

The quantities~(\ref{normsLS}) are norms, and they can be used to define a 
countably normed topology (i.e., a meaning of sequence convergence)
$\tau _{\mathbf \Phi _{\rm exp}}$ on $\mathbf \Phi _{\rm exp}$:  
\begin{equation}
      \varphi ^{\pm}_{\alpha}\, 
       \mapupdown{\tau_{\mathbf \Phi _{\rm exp}}}{\alpha \to \infty}
      \, \varphi ^{\pm} \quad {\rm iff} \quad  
      \| \varphi ^{\pm}_{\alpha}-\varphi ^{\pm} \| _{n,n'} 
      \, \mapupdown{}{\alpha \to \infty}\, 0 \, , \quad n,n'=0,1,2, \ldots 
            \, . 
\end{equation}   

Once we have constructed the space $\mathbf \Phi _{\rm exp}$, we can 
construct its dual $\mathbf \Phi _{\rm exp}^{\prime}$ and antidual
$\mathbf \Phi _{\rm exp}^{\times}$ spaces as the 
spaces of, respectively, linear and antilinear continuous 
functionals over $\mathbf \Phi _{\rm exp}$, and therewith 
the rigged Hilbert spaces~(\ref{rhsexpp}) and (\ref{rhsexpt}). The 
Lippmann-Schwinger bras and kets are, respectively,
linear and antilinear continuous functionals over ${\mathbf \Phi}_{\rm exp}$, 
i.e., $\langle ^{\pm}q| \in {\mathbf \Phi}_{\rm exp}^{\prime}$ and  
$|q^{\pm}\rangle \in {\mathbf \Phi}_{\rm exp}^{\times}$. As well,
$\langle ^{\pm}q|$ and $|q^{\pm}\rangle$ are, respectively, 
``left'' and ``right'' eigenvectors of $H$ with eigenvalue 
$\hbar ^2/(2m) \, q^2$. 

The following proposition, whose proof can be found in 
appendix~\ref{sec:A3}, encapsulates the results of this section:

\vskip0.5cm

\newtheorem*{Prop2}{Proposition~2}
\begin{Prop2} \label{Prop2} The triplets of spaces~(\ref{rhsexpp}) and 
(\ref{rhsexpt}) are rigged Hilbert spaces, and they satisfy all
the requirements to accommodate the analytic continuation of the
Lippmann-Schwinger bras and kets. More specifically,
\begin{itemize}
\item[({\it i})] The $\| \cdot \|_{n,n'}$ are norms.
\item[({\it ii})] The space ${\mathbf \Phi}_{\rm exp}$ is dense in
$L^2([0,\infty ),\rmd r)$.
\item[({\it iii})] The space ${\mathbf \Phi}_{\rm exp}$ is invariant under
the action of the Hamiltonian, and $H$ is 
${\mathbf \Phi _{\rm exp}}$-continuous.
\item[({\it iv})] The kets $|q^{\pm}\rangle$ are continuous, 
{\it antilinear} functionals over ${\mathbf \Phi}_{\rm exp}$, i.e., 
$|q^{\pm}\rangle  \in {\mathbf \Phi}_{\rm exp}^{\times}$.
\item[({\it v})] The kets $|q^{\pm}\rangle$ are ``right'' eigenvectors 
of $H$ with eigenvalue $\frac{\hbar ^2}{2m}q^2$:
\numparts
\begin{equation}
      H|q^{\pm}\rangle= \frac{\hbar ^2}{2m}q^2 \, |q^{\pm}\rangle \, ;
       \label{keigeeqa}
\end{equation}
that is,
\begin{equation}
      \langle \varphi ^{\pm}|H|q^{\pm}\rangle =
     \frac{\hbar ^2}{2m}q^2 \langle \varphi ^{\pm}|H|q^{\pm}\rangle \, , \quad 
       \varphi ^{\pm} \in {\mathbf \Phi}_{\rm exp}  \, .
      \label{keigeeqbis} 
\end{equation}
\endnumparts
\item[({\it vi})] The bras $\langle ^{\pm}q|$ are continuous, 
{\it linear} functionals over ${\mathbf \Phi}_{\rm exp}$, i.e., 
$\langle ^{\pm}q| \in {\mathbf \Phi}_{\rm exp}^{\prime}$.
\item[({\it vii})]The bras $\langle ^{\pm}q|$ are ``left'' 
eigenvectors of $H$ with eigenvalue $\frac{\hbar ^2}{2m} q^2$:
\numparts
\begin{equation}
      \langle ^{\pm}q|H=\frac{\hbar ^2}{2m} q^2  \langle ^{\pm}q|\, ;
             \label{kpssleftkeofHa}
\end{equation}
that is,
\begin{equation}
       \langle ^{\pm}q|H|\varphi ^{\pm}\rangle 
       = \frac{\hbar ^2}{2m}q^2 \langle ^{\pm}q| \varphi ^{\pm}\rangle \, .
        \label{kpssleftkeofHb}   
\end{equation}
\endnumparts
\end{itemize}
\end{Prop2}

\vskip0.5cm

Equations~(\ref{keigeeqa}) and (\ref{kpssleftkeofHa}) can be rewritten in 
terms of the complex energy $z$ as
\begin{equation}
       H|z^{\pm}\rangle=z|z^{\pm}\rangle \, ,
         \label{eigenzi}
\end{equation}
\begin{equation}
      \langle ^{\pm}z|H = z \langle ^{\pm}z| \, .
           \label{eigenzibra}
\end{equation} 
Note that the bra eigenequation~(\ref{eigenzibra}) is not given by
$\langle ^{\pm}z|H = \overline{z}\langle ^{\pm}z|$, as one may naively
expect from formally obtaining~(\ref{eigenzibra}) by Hermitian conjugation
of the ket eigenequation~(\ref{eigenzi}). The reason
lies in that the function $\overline{z}$ is not analytic, so when
one obtains the bra eigenequation by Hermitian conjugation of the ket
eigenequation, one has to use $\overline{\overline{z}}=z$. The following
chain of equalities further clarifies this point:
\begin{equation}
      \langle ^{\pm}z|H|\varphi ^{\pm}\rangle = 
       z\langle ^{\pm}z|\varphi ^{\pm}\rangle =
      z \overline{\langle \varphi^{\pm}|\overline{z}^{\pm}\rangle} =
     \overline{\overline{z}\langle \varphi^{\pm}|\overline{z}^{\pm}\rangle} =
   \overline{\langle \varphi^{\pm}|H|\overline{z}^{\pm}\rangle} \, .
\end{equation}    

The ``free'' bras~(\ref{anconfbra}) and kets~(\ref{anconfket}) can also be 
accommodated within the rigged Hilbert spaces~(\ref{rhsexpp}) and   
(\ref{rhsexpt}). To see this, one just has to recall the 
estimate~(\ref{estchiozero}). One can then show, in complete analogy with
the Lippmann-Schwinger bras and kets, that $\langle q|$ belongs to
${\mathbf \Phi}_{\rm exp}^{\prime}$, and that $|q\rangle$ belongs to
${\mathbf \Phi}_{\rm exp}^{\times}$. As well, one can easily prove that
$\langle q|$ and $|q\rangle$ are, respectively, ``left'' and ``right''
eigenvectors of $H_0$ with eigenvalue $\frac{\hbar ^2}{2m}q^2$.

It is clear that there is a 1:1 correspondence between bras and kets
also when the energy and the wave number become complex. The following
table summarizes such correspondence:
\begin{equation}
\begin{tabular}{|c||ccc|c}
    \hline
    \quad & wave number &  $\longleftrightarrow$   &  energy  \\
     \hline 
       bra  &  $\langle ^{\pm}q|$, \  $\langle q|$ 
           &  $\longleftrightarrow$ &    
               $\langle ^{\pm}z|$, \  $\langle z|$ \\ [1ex]
     $\updownarrow$   &  $\updownarrow$ & \quad & $\updownarrow$ \\ [1ex]
        ket  &  $|q^{\pm}\rangle$, \  $|q\rangle$  & 
           $\longleftrightarrow$  &
                  $|z^{\pm}\rangle$, \ $|z\rangle$  \\
        \hline
\end{tabular}
\end{equation}

\section{The wave number representations of the rigged Hilbert spaces, bras
and kets}
\setcounter{equation}{0}
\label{sec:errHsanaly}

We turn now to obtain and characterize the wave number representations of
the rigged Hilbert spaces~(\ref{rhsexpp}) and (\ref{rhsexpt}) as well as of
the ``in'' and ``out'' wave functions, bras and kets. The wave number 
representations are very useful, because sometimes they differentiate between
the ``in'' and the ``out'' boundary conditions in a more clear way than
the position representation.

\subsection{The wave number representations of the rigged Hilbert spaces}

The ``in'' ($+$) and the ``out'' ($-$) wave number representations of 
${\mathbf \Phi}_{\rm exp}$ are readily obtained by means of the unitary 
operators ${\cal F}_{\pm}$ of Eq.~(\ref{F+-slthe}):
\begin{equation}
     {\cal F}_{\pm}{\mathbf \Phi}_{\rm exp} \equiv  
          \widehat{\mathbf \Phi}_{\pm {\rm exp}} \, ,
\end{equation}
which in turn yield the wave number representations of the rigged Hilbert
spaces~(\ref{rhsexpp}) and (\ref{rhsexpt}):
\numparts
\begin{equation}
     \widehat{\mathbf \Phi}_{\pm {\rm exp}} \subset L^2([0,\infty ),\rmd k)
      \subset \widehat{\mathbf \Phi}_{\pm {\rm exp}}^{\prime}   \, ,
\end{equation}
\begin{equation}
     \widehat{\mathbf \Phi}_{\pm {\rm exp}} \subset L^2([0,\infty ),\rmd k)
      \subset \widehat{\mathbf \Phi}_{\pm {\rm exp}}^{\times}   \, .
\end{equation}
\endnumparts

The functions $\widehat{\varphi}^{\pm}(q)$ in 
$\widehat{\mathbf \Phi}_{\pm {\rm exp}}$ are obviously the analytic 
continuation of $\widehat{\varphi}^{\pm}(k)$ from the positive $k$-axis into
the whole $k$-plane. One can easily show that
\begin{equation}
      \widehat{\varphi}^{\pm}(q)= \langle ^{\pm}q|\varphi ^{\pm}\rangle \, , 
           \label{expreofbraintwf}
\end{equation}
and that
\begin{equation}
      \overline{\widehat{\varphi}^{\pm}(\overline{q})}=
            \langle \varphi ^{\pm}|q^{\pm} \rangle  \, .
     \label{expreofketintwf}
\end{equation}
The poles of the Lippmann-Schwinger eigenfunctions are carried over into the
analytic continuation of the wave functions: The function
$\widehat{\varphi}^{\pm}(q)$ is analytic everywhere except at
$Z_{\mp}$, where its poles are located, and 
$\overline{\widehat{\varphi}^{\pm}(\overline{q})}$ is analytic everywhere 
except at $Z_{\pm}$, where its poles are located.

That $\widehat{\varphi}^{\pm}(k)$ can be analytically 
continued into $\widehat{\varphi}^{\pm}(q)$ is made possible by the falloff of 
$\varphi ^{\pm}(r)$ at infinity. The falloff of $\varphi ^{\pm}(r)$
also limits the growth of $\widehat{\varphi}^{\pm}(q)$. Such growth is 
provided by the following proposition:

\vskip0.5cm

\newtheorem*{Prop3}{Proposition~3}
\begin{Prop3} In the lower half of the $k$-plane, $\widehat{\varphi}^+(q)$ 
grows slower than $\rme ^{|{\rm Im}(q)|^2}$. More precisely, for every 
positive integer $n'$, and for each $\alpha >0$, the following estimate holds:
\begin{equation}
    | (1+ \frac{\hbar ^2}{2m}q^2)^{n'} \widehat{\varphi}^+(q)| \leq 
         C \,  \rme ^{\frac{\, \,  |{\rm Im}(q)|^2}{2\alpha }} \, ,
    \quad    {\rm Im}(q)\leq 0   \, ,
    \label{boundinwhvalp+}
\end{equation} 
where the constant $C$ depends on ${n'}$, ${\varphi}^+$ and $\alpha$, but not
on $q$. In the upper half plane, $\widehat{\varphi}^{+}(q)$ is infinity 
whenever $q \in Z_{-}$. As $|q|$ tends to $\infty$ in the upper half plane,
it holds that
\begin{equation}
    | (1+ \frac{\hbar ^2}{2m}q^2)^{n'} \widehat{\varphi}^+(q)| \leq 
         C \,\frac{1}{|\lambda (-q)|}  
              \rme ^{\frac{\, \,  |{\rm Im}(q)|^2}{2\alpha }} 
            \, , \quad
            (|q|\to \infty \, , \  {\rm Im}(q)>0) \, ,
    \label{boundinwhvaup+}
\end{equation}
where $\lambda (-q)$ is given by Proposition~1.

The above estimates are satisfied by $\widehat{\varphi}^-(q)$ when we
exchange the upper for the lower half plane:
\begin{equation}
    |(1+ \frac{\hbar ^2}{2m}q^2)^{n'} \widehat{\varphi}^-(q)|\leq 
               C \,  \rme ^{\frac{\, \, |{\rm Im}(q)|^2}{2\alpha}} \, ,
    \quad    {\rm Im}(q)\geq 0   \, .
      \label{boundinwhphp-}
\end{equation} 
\begin{equation}
    | (1+ \frac{\hbar ^2}{2m}q^2)^{n'} \widehat{\varphi}^-(q)| \leq 
         C \,\frac{1}{|\lambda (q)|}  
              \rme ^{\frac{\, \,  |{\rm Im}(q)|^2}{2\alpha }} 
            \, , \quad
            (|q|\to \infty \, , \  {\rm Im}(q)<0) \, ,
    \label{boundinwhvaup-}
\end{equation}

\end{Prop3}

\vskip0.5cm

The proof of Proposition~3 can be found in appendix~\ref{sec:A3}, and it
is based on the theory of $M$ and $\Omega$ 
functions, see Ref.~\cite{GELFANDIII} and appendix~\ref{sec:omegafunc}. For
our purposes, the most important result is
\begin{equation}
        xy \leq \frac{x^n}{n} + \frac{y^{n'}}{n'} \, ,
       \label{xylexpyq}
\end{equation}
where $x,y \geq 0$ and
\begin{equation}
        \frac{1}{n} + \frac{1}{n'} = 1 \, . 
\end{equation}
Equation~(\ref{xylexpyq}) can be used to show that when $\varphi ^{\pm}(r)$ 
falls off faster than $\rme ^{-r^n}$, then, away from its poles, 
$\widehat{\varphi}^{\pm}(q)$ grows slower than $\rme ^{|{\rm Im}(q)|^{n'}}$. In
this paper, we use $n=n'=2$.

The bounds in Proposition~3 are very wasteful when $|q|\to 0$, where 
$\widehat{\varphi}^{\pm}(q)$ actually tends to $0$. This happened because in 
the proof of Proposition~3, we dismiss the factor $|q|r/(1+|q|r)$. Dismissing
this factor should not be the cause of concern, since the most crucial
behavior of $\widehat{\varphi}^{\pm}(q)$ occurs in the limit $|q|\to \infty$. 

It is interesting to compare the growth of our test functions with the
growth of the test functions used by Bollini 
{\it et al.}~\cite{BOLLINI1,BOLLINI2}. In~\cite{BOLLINI1,BOLLINI2}, 
$\varphi (r)$ falls off like $\rme ^{-r}$, and 
therefore $|\widehat{\varphi}(p)|$ grows faster 
than any exponential of $|{\rm Im}(p)|^n$, where $p$ denotes the complex 
momentum and $n$ can be any positive integer. In the present paper, 
$\varphi (r)$ falls off like $\rme ^{-r^2}$, and therefore 
$|\widehat{\varphi}^{\pm}(q)|$ grows like $\rme ^{|{\rm Im}(q)|^2}$ away from
its poles.

It is also interesting to compare our approach with that based on Hardy 
functions~\cite{GADELLA84,BG,GADELLA-ORDONEZ,BRTK,CAGADA,GAGO,DIS}. From
Eq.~(\ref{wnresifunc}), one can obtain the analytic and growth properties
of the wave functions in the energy representation, 
$\widehat{\varphi}^{\pm}(z)$, from those of 
$\widehat{\varphi}^{\pm}(q)$. Since by Proposition~3 the wave functions
$\widehat{\varphi}^{\pm}(q)$ blow up exponentially in the infinity arc
of the wave number plane, the wave functions $\widehat{\varphi}^{\pm}(z)$
also blow up exponentially in the infinity arcs of the Riemann 
surface. Therefore, $\widehat{\varphi}^{\pm}(z)$ are not Hardy functions, 
because if they were, they would tend to zero in one of the infinite semi-arcs 
of the Riemann surface. Hence, our approach is different from that based 
on Hardy functions.

\subsection{The wave number representation of the Lippmann-Schwinger bras
and kets}

The wave number representation of the bras $\langle ^{\pm}q|$ and kets 
$|q^{\pm}\rangle$ is defined as
\begin{equation}
     \langle ^{\pm}\widehat{q}| \equiv  \langle ^{\pm}q|{\cal F}_{\pm} \, , 
        \label{pmqbrsinwnr}
\end{equation}
\begin{equation}
      |\widehat{q}^{\pm}\rangle \equiv {\cal F}_{\pm}|q^{\pm}\rangle  \, .
         \label{pmqketsinwnr}
\end{equation}

The bras $\langle ^{\pm}q|$ and kets $|q^{\pm}\rangle$ are obviously different
from their wave-number representations $\langle ^{\pm}\widehat{q}|$ and
$|\widehat{q}^{\pm}\rangle$, and such difference can be better understood 
through a simpler example. Consider the 1D momentum operator 
$P=-\rmi \hbar \rmd /\rmd x$. In the position representation, the 
$\delta$-normalized eigenfunctions of $P$ are the exponentials
$\frac{1}{\sqrt{2\pi \hbar}}\rme ^{\rmi px/\hbar}$, and these are the
analog of $|q^{\pm}\rangle$. In the momentum representation, which is 
obtained by Fourier transforming the position representation, the 
eigenfunctions of the momentum operator become the delta function 
$\delta (p-p')$, and these are the analog of $|\widehat{q}^{\pm}\rangle$.

When $q$ does not belong to $Z_{\mp}$, the bras $\langle ^{\pm}\widehat{q}|$ 
act as the linear complex delta functional, as the following chain of 
equalities show:
\begin{eqnarray}
       \langle ^{\pm}\widehat{q}| \widehat{\varphi}^{\pm} \rangle &=
   \langle ^{\pm}q|{\cal F}_{\pm}|\widehat{\varphi}^{\pm} \rangle  
    &  \hskip1cm \mbox{by (\ref{pmqbrsinwnr})}  \nonumber \\
    &=\langle ^{\pm}q|{\cal F}_{\pm}^{\dagger}\widehat{\varphi}^{\pm} \rangle  
         &  \hskip1cm \mbox{by (\ref{beigenrhsp})}           
   \nonumber \\
   &= \langle ^{\pm}q|{\varphi}^{\pm} \rangle  
            \nonumber \\
     &= \widehat{\varphi}^{\pm}(q) \, , \quad q \notin  Z_{\mp}  
    &  \hskip1cm \mbox{by (\ref{expreofbraintwf})} \, .   
         \label{hatpmbrnp}
\end{eqnarray}
When $q$ belongs to $Z_{\mp}$, the wave function $\widehat{\varphi}^{\pm}(q)$
has a pole at $q$, and therefore the bra $\langle ^{\pm}\widehat{q}|$ acts
as the linear residue functional:
\begin{equation}
       \langle ^{\pm}\widehat{q}| \widehat{\varphi}^{\pm} \rangle =
    {\rm res}  \, [\widehat{\varphi}^{\pm}(q) ] \, , 
         \quad q \in  Z_{\mp} \, .   
    \label{abraatporesi}
\end{equation}

Similarly, when $q$ does not belong to $Z_{\pm}$, the kets $|q^{\pm}\rangle$
act as the antilinear complex delta functional, as the following chain of 
equalities show:
\begin{eqnarray}
    \langle \widehat{\varphi}^{\pm}| \widehat{q}^{\pm} \rangle &=
    \langle \widehat{\varphi}^{\pm}|{\cal F}_{\pm}|q^{\pm} \rangle 
      & \hskip1cm \mbox{by (\ref{pmqketsinwnr})} \nonumber \\
   &= \langle {\cal F}_{\pm}^{\dagger}\widehat{\varphi}^{\pm}|q^{\pm} \rangle 
        &  \hskip1cm \mbox{by (\ref{keigenrhsc})}  \nonumber \\
     &= \langle \varphi ^{\pm}|q^{\pm} \rangle \nonumber \\
    &= \overline{\widehat{\varphi}^{\pm}(\overline{q})} \, , 
             \quad q \notin  Z_{\pm} 
     &   \hskip1cm  \mbox{by (\ref{expreofketintwf})} \, .   
         \label{whqmpkw}
\end{eqnarray}
When $q$ belongs to $Z_{\pm}$, the wave function 
$\overline{\widehat{\varphi}^{\pm}(\overline{q})}$
has a pole at $q$, and therefore the ket $|q^{\pm}\rangle$ acts
as the antilinear residue functional:
\begin{equation}
       \langle \widehat{\varphi}^{\pm}| \widehat{q}^{\pm} \rangle =
    {\rm res} \, [ \overline{\widehat{\varphi}^{\pm}(\overline{q})} ] 
            \, ,  \quad q \in  Z_{\pm} \, .   
         \label{ketapoleresi}
\end{equation}

The complex delta functional and the residue functional can be written
in more familiar terms as follows. By using the resolution of the 
identity~(\ref{residenpm}), we can formally write the action of 
$\langle ^{\pm}\widehat{q}|$ as an integral operator and obtain
\begin{eqnarray}
     \langle ^{\pm}\widehat{q}| \widehat{\varphi}^{\pm} \rangle &=     
      \langle ^{\pm}q|{\varphi}^{\pm} \rangle  \nonumber \\
     & = \int_0^{\infty}\rmd k \, 
      \langle ^{\pm}q|k^{\pm}\rangle \langle ^{\pm}k|{\varphi}^{\pm} \rangle
     \nonumber \\
      & = \int_0^{\infty}\rmd k \, 
      \langle ^{\pm}q|k^{\pm}\rangle \, \widehat{\varphi}^{\pm} (k)  \, .
        \label{braqpmintope}
\end{eqnarray}
Comparison of (\ref{braqpmintope}) with (\ref{hatpmbrnp}) shows
that when $q \notin Z_{\mp}$, $\langle ^{\pm}q|k^{\pm}\rangle$ coincides
with the complex delta function at $q$:
\begin{equation}
       \langle ^{\pm}q|k^{\pm}\rangle = \delta (k-q) \, , \quad 
        q \notin Z_{\mp} \, .
           \label{comdpbrale}
\end{equation}
Note that when $q$ is positive, Eq.~(\ref{comdpbrale}) reduces to the 
standard $\delta$-function normalization. When $q \in Z_{\mp}$, comparison 
of (\ref{braqpmintope}) with 
(\ref{abraatporesi}) implies that $\langle ^{\pm}q|k^{\pm}\rangle$ coincides
with the residue distribution at $q$:
\begin{equation}
       \langle ^{\pm}q|k^{\pm}\rangle = {\rm res}  \, [\, \cdot \, ]_q \, , 
    \quad  q \in Z_{\mp} \, .
\end{equation}

Similarly, by using~(\ref{residenpm}) we can formally write the action of 
$|\widehat{q}^{\pm}\rangle$ as an integral operator:
\begin{eqnarray}
     \langle  \widehat{\varphi}^{\pm}|\widehat{q}^{\pm} \rangle &=     
      \langle {\varphi}^{\pm}|q^{\pm} \rangle  \nonumber \\
     & = \int_0^{\infty}\rmd k \, 
         \langle {\varphi}^{\pm}|k^{\pm}\rangle \langle ^{\pm}k|q^{\pm} \rangle
     \nonumber \\
      & = \int_0^{\infty}\rmd k \, 
      \overline{\varphi ^{\pm}(k)} \langle ^{\pm}k|q^{\pm} \rangle \, .
        \label{acwpmq}
\end{eqnarray}   
By comparing (\ref{acwpmq}) with (\ref{whqmpkw}), we deduce that when 
$q \notin Z_{\pm}$, $\langle ^{\pm}k|q^{\pm}\rangle$ coincides with
the complex delta function at $q$:
\begin{equation}
       \langle ^{\pm}k|q^{\pm}\rangle = \delta (k-q) \, , \quad 
        q \notin Z_{\pm} \, .
\end{equation}
When $q \in Z_{\pm}$, comparison of (\ref{acwpmq}) with (\ref{ketapoleresi}) 
lead us to identify $\langle ^{\pm}k|q^{\pm}\rangle$ as
the residue distribution at $q$:
\begin{equation}
       \langle ^{\pm}k|q^{\pm}\rangle = {\rm res}  \, [\, \cdot \, ]_q \, , 
    \quad  q \in Z_{\pm} \, .
\end{equation}

It is important to note that, with a given test function, the complex delta 
function and the residue distribution at $q$ associate, respectively, the value
and the residue of the analytic continuation of the test function at 
$q$. This
is why when those distributions act on $\overline{\widehat{\varphi}^{\pm}(k)}$
as in Eq.~(\ref{acwpmq}), the final result is respectively 
$\overline{\widehat{\varphi}^{\pm}(\overline{q})}$ and
${\rm res} \, [ \overline{\widehat{\varphi}^{\pm}(\overline{q})} ]$, rather
than $\overline{\widehat{\varphi}^{\pm}(q)}$ and
${\rm res} \, [ \overline{\widehat{\varphi}^{\pm}(q)} ]$, since 
the analytic continuation of $\overline{\widehat{\varphi}^{\pm}(k)}$ is
$\overline{\widehat{\varphi}^{\pm}(\overline{q})}$ rather than
$\overline{\widehat{\varphi}^{\pm}(q)}$.

\subsection{The ``free'' wave number representation}

One can also construct the wave number representation associated with
the ``free'' Hamiltonian. Since its construction follows the same steps
as that of the ``in'' and ``out'' wave number representations, we shall
simply list the main results.

The unitary operator ${\cal F}_{0}$ in Eq.~(\ref{F0slthe}) provides the 
``free'' wave number representation of the space of test functions:
\begin{equation}
     {\cal F}_{0}{\mathbf \Phi}_{\rm exp} \equiv  
          \widehat{\mathbf \Phi}_{0{\rm exp}} \, ,
\end{equation}
which in turn yields the ``free'' wave number representation of the rigged 
Hilbert spaces~(\ref{rhsexpp}) and (\ref{rhsexpt}):
\numparts
\begin{equation}
     \widehat{\mathbf \Phi}_{0{\rm exp}} \subset L^2([0,\infty ),\rmd k)
      \subset \widehat{\mathbf \Phi}_{0{\rm exp}}^{\prime}   \, ,
\end{equation}
\begin{equation}
     \widehat{\mathbf \Phi}_{0{\rm exp}} \subset L^2([0,\infty ),\rmd k)
      \subset \widehat{\mathbf \Phi}_{0{\rm exp}}^{\times}   \, .
\end{equation}
\endnumparts

The functions $\widehat{\varphi}(q)$ in 
$\widehat{\mathbf \Phi}_{0{\rm exp}}$ are the analytic 
continuation of $\widehat{\varphi}(k)$ from the positive $k$-axis into
the whole $k$-plane. One can easily show that
\begin{equation}
      \widehat{\varphi}(q)= \langle q|\varphi \rangle \, , \quad
          q \in {\mathbb C} \, .
           \label{expreofbraintwf0}
\end{equation}
and that
\begin{equation}
      \overline{\widehat{\varphi}(\overline{q})}=
            \langle \varphi |q\rangle \, , \quad 
       q \in {\mathbb C}  \, .
     \label{expreofketintwf0}
\end{equation}
The functions $\widehat{\varphi}(q)$ are analytic in the whole $k$-plane, and 
they satisfy the following estimate for any $\alpha >0$ and for any positive 
integer $n'$:
\begin{equation}
    | (1+ \frac{\hbar ^2}{2m}q^2)^{n'} \widehat{\varphi}(q)| \leq 
         C \,  \rme ^{\frac{\, \,  |{\rm Im}(q)|^2}{2\alpha }} \, ,
    \quad    q \in {\mathbb C}  \, ,
    \label{boundinwhvalp0}
\end{equation} 
where the constant $C$ depends on ${n'}$, ${\varphi}$ and $\alpha$, but not
on $q$. 

The ``free'' wave number representation of $\langle q|$ and $|q\rangle$ is 
defined as
\begin{equation}
     \langle \widehat{q}| \equiv  \langle q|{\cal F}_{0} \, , 
        \label{pmqbrsinwnr0}
\end{equation}
\begin{equation}
      |\widehat{q}\rangle \equiv {\cal F}_{0}|q\rangle  \, .
         \label{pmqketsinwnr0}
\end{equation}
One can easily show that $\langle \widehat{q}|$ and $|\widehat{q}\rangle$ are,
respectively, the linear and antilinear complex delta functionals.

\section{The time evolution of the analytic continuation of the 
Lippmann-Schwinger bras and kets}
\setcounter{equation}{0}
\label{sec:timeevolu}

In Ref.~\cite{LS1}, we obtained the time evolution of the ``in,'' as well
as of the ``out,'' wave functions, bras and kets. In terms of the wave 
number, the time evolution of the wave functions $\varphi ^{\pm}$ is given by
\begin{equation}
     \varphi ^{\pm} (r;t)=   
      \left( \rme^{-\rmi Ht/\hbar}\varphi ^{\pm}\right) (r)=
       \int_0^{\infty} \rmd k\, \rme ^{-\rmi k^2 \hbar t/(2m)}
       \widehat{\varphi}^{\pm}(k) \chi ^{\pm}(r;k)  \, ,
       \label{timevownwf} 
\end{equation}
which is valid for $-\infty < t < \infty$. Equation~(\ref{timevownwf}) is 
equivalent to saying that the
operator $\rme ^{-\rmi Ht/\hbar}$ acts, in the wave number representation, as
multiplication by $\rme ^{-\rmi k^2 \hbar t/(2m)}$:
\begin{equation}
     \widehat{\varphi}^{\pm}(k;t)=
    \left( \rme ^{-\rmi \widehat{H}t/\hbar}\widehat{\varphi}^{\pm}\right)(k)=
      \rme ^{-\rmi k^2 \hbar t/(2m)} \widehat{\varphi}^{\pm}(k) \, . 
      \label{timevolwispwnare}
\end{equation}
For $k$ positive, the time evolution of the Lippmann-Schwinger bras and kets 
is given by
\begin{equation}
      \langle ^{\pm}k|\rme ^{-\rmi Ht/\hbar} =  \rme ^{\rmi k^2 \hbar t/(2m)}
       \langle ^{\pm}k|  \, , 
\end{equation}
\begin{equation}
      \rme ^{-\rmi Ht/\hbar}|k^{\pm}\rangle =  \rme ^{-\rmi k^2 \hbar t/(2m)}
        |k^{\pm}\rangle  \, .
\end{equation}

In this section, we analytically continue the above equations into the
$k$-plane, thereby obtaining the time evolution of the
analytic continuation of the ``in,'' as well as of the ``out,'' wave functions,
bras and kets. As we shall see, such continuation entails the imposition of 
a time asymmetric boundary condition upon the time evolution.

\subsection{The analytic continuation of the time evolution}

The analytic continuation of Eq.~(\ref{timevolwispwnare}) is given by
\begin{equation}
    \widehat{\varphi}^{\pm}(q;t)=
    \left( \rme ^{-\rmi \widehat{H}t/\hbar}\widehat{\varphi}^{\pm}\right)(q)=
       \rme ^{-\rmi q^2 \hbar t/(2m)} \widehat{\varphi}^{\pm}(q) \, .
      \label{timevolwispwnareacon}
\end{equation}
The factor $\rme ^{-\rmi q^2 \hbar t/(2m)}$ does not change the
analytic properties of $\widehat{\varphi}^{\pm}(q)$. It does, however, change
the growth properties of $\widehat{\varphi}^{\pm}(q)$ depending on the 
sign of $t$ and on the quadrant of the complex plane. As can be easily seen,
\begin{equation}
   \rme ^{-\rmi q^2 \hbar t/(2m)} \mapupdown{\quad}{\ |q|\to \infty \ } 0 \, ,
         \begin{array}{c}
            \quad  t>0 \, , \ q\in {\rm 2nd, 4th},\\  
                            \hskip-0.4cm  {\rm or} \\
            \quad  t<0 \, ,  \ q\in {\rm 1st, 3rd},
         \end{array}
          \label{timelimit0} 
\end{equation}
\begin{equation}
  \rme ^{-\rmi q^2\hbar t/(2m)}\mapupdown{\quad}{\ |q|\to \infty \ }\infty \, ,
         \begin{array}{c}     
                  \quad t<0 \, , \ q\in {\rm 2nd, 4th}, \\ 
                                    \hskip-0.4cm  {\rm or} \\
                  \quad t>0 \, , \  q\in {\rm 1st, 3rd},
         \end{array}
       \label{timelimitinfty} 
\end{equation}
where 1st, 2nd, 3rd and 4th denote, respectively, the first, second, third
and fourth quadrants of the $k$-plane. Thus, even though 
$\widehat{\varphi}^{\pm}(q)$ blows up exponentially for large $q$, 
$\widehat{\varphi}^{\pm}(q;t)$ goes to zero in the infinite arc of the second 
and fourth quadrants when $t>0$. In the infinite arc of the first and third 
quadrants, $\widehat{\varphi}^{\pm}(q;t)$ goes to zero when $t<0$. Hence, the 
analytic continuation of the time evolution changes the growth properties 
of the wave functions and introduces a time asymmetry. 

\begin{figure}[ht]
 \begin{center}
   \includegraphics{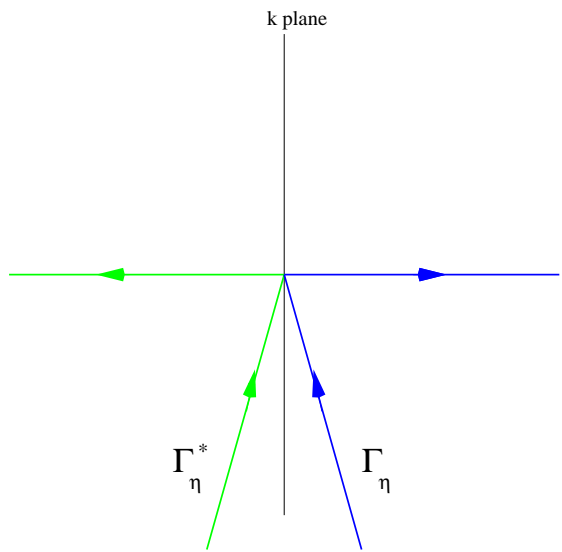}
    \caption{The contours $\Gamma _{\eta}$ and $\Gamma _{\eta}^*$. The 
straight lines in the third and fourth quadrants form an angle $\eta$ with
the negative imaginary axis, $\eta$ being infinitesimally small.}
    \label{fig:etas}
 \end{center}
\end{figure}

In practical situations, the importance of the limits~(\ref{timelimit0}) lies 
in the fact that they enable us to continue certain contour integrals all the 
way to the infinite arc of a quadrant in such a way that such infinite arc 
does not contribute to the integral. For example, if $\Gamma _{\eta}$ and 
$\Gamma _{\eta}^*$ denote the contours depicted in Fig.~\ref{fig:etas}, then 
Cauchy's theorem and the bound~(\ref{boundinwhvalp+}), together with the 
limits~(\ref{timelimit0}), yield
\numparts 
\begin{equation}
     \int_{\Gamma_{\eta}} \rmd q \, \rme ^{-\rmi q^2 \hbar t/(2m)}
          \widehat{\varphi}^+(q) = 0 \, ,
         \quad t>0 \, ,
\end{equation}
\begin{equation}
     \int_{\Gamma_{\eta}^*} \rmd q \, \rme ^{-\rmi q^2 \hbar t/(2m)}
         \widehat{\varphi}^+(q) = 0 \, , 
      \quad t<0 \, .
\end{equation}
\endnumparts
These two equations exemplify the different behavior of 
$\widehat{\varphi}^+(q;t)$ in different quadrants of the $k$-plane for 
opposite signs of time.

Our next objective is to analytically continue Eq.~(\ref{timevownwf}). In
order to do so, we define the contour $\gamma _\varepsilon$ as the radial
path in the fourth quadrant that forms an angle $-\varepsilon$ with the 
positive $k$-axis, see Fig.~\ref{fig:gammas}a. Then,
\begin{equation}
     \varphi ^{\pm}(r;t)= 
       \int_{\gamma _\varepsilon} \rmd q\, \rme ^{-\rmi q^2 \hbar t/(2m)}
       \widehat{\varphi}^{\pm}(q) \chi ^{\pm}(r;q)  \, .
       \label{timevownwfAC} 
\end{equation}
Because by~(\ref{timelimit0}) and (\ref{timelimitinfty}) 
$\rme ^{-\rmi q^2 \hbar t/(2m)}$ tends to zero in the infinite arc
of the fourth quadrant only for positive times, the time 
evolution~(\ref{timevownwfAC}) is defined only for $t>0$. Thus, the analytic 
continuation into the fourth quadrant converts the time evolution group 
$\rme ^{-\rmi Ht/\hbar}$ into a semigroup. We shall denote this semigroup 
by $\rme _+^{-\rmi Ht/\hbar}$:
\begin{equation}
    \hskip-1cm   \varphi ^{\pm}(r;t)= 
        \left( \rme _+^{-\rmi Ht/\hbar}\varphi ^{\pm}\right) (r)=
       \int_{\gamma _\varepsilon} \rmd q\, \rme ^{-\rmi q^2 \hbar t/(2m)}
       \widehat{\varphi}^{\pm}(q) \chi ^{\pm}(r;q)  \, , \quad t>0 \, .
       \label{timevownwfAC+se} 
\end{equation}
Similarly, because by~(\ref{timelimit0}) and (\ref{timelimitinfty}) 
$\rme ^{-\rmi q^2 \hbar t/(2m)}$ tends to zero in the infinite arc of the
third quadrant only for negative times, the analytic continuation of the
time evolution into the 3rd quadrant converts $\rme ^{-\rmi Ht/\hbar}$ into
a semigroup valid for $t<0$ only. We shall denote this semigroup by 
$\rme _-^{-\rmi Ht/\hbar}$:
\begin{equation}
     \hskip-1.5cm   \varphi ^{\pm}(r;t)= 
      \left( \rme _-^{-\rmi Ht/\hbar}\varphi ^{\pm}\right) (r)= -
       \int_{\gamma _\varepsilon ^*} \rmd q\, \rme ^{-\rmi q^2 \hbar t/(2m)}
       \overline{\widehat{\varphi}^{\pm}(\overline{q})} 
        \  \overline{\chi ^{\pm}(r;\overline{q})}  \, , \quad t<0 \, ,
       \label{timevownwfAC-se} 
\end{equation}
where $\gamma_\varepsilon ^*$ is the mirror image of $\gamma _{\varepsilon}$
with respect to the imaginary axis, see Fig.~\ref{fig:gammas}a. In 
Eqs.~(\ref{timevownwfAC+se}) and (\ref{timevownwfAC-se}), $\varepsilon$ is 
small enough so that $\gamma _{\varepsilon}$ and $\gamma _{\varepsilon}^*$ do 
not pick up resonance contributions. (If necessary to avoid resonances, the 
contours $\gamma _{\varepsilon}$ and $\gamma _{\varepsilon}^*$ may be bent.)

Note that the analogous analytic continuation into the first 
quadrant yields a semigroup for $t<0$, whereas the continuation into
the second quadrant yields a semigroup for $t>0$. Note also the similarity
of these analytic continuations with the $\pm \rmi \varepsilon$ prescriptions. 

By comparing the semigroup evolution,
\begin{equation}
     \varphi ^{\pm}(r;t) = \rme _+^{-\rmi Ht/\hbar }\varphi ^{\pm}(r) \, , 
              \quad  t>0 \ {\rm only} \, , 
\end{equation}
with the standard time evolution,
\begin{equation}
     \varphi ^{\pm}(r;t) = \rme^{-\rmi Ht/\hbar }\varphi ^{\pm}(r) \, ,
        \quad  t\in {\mathbb R} \, ,
\end{equation}
we are able to conclude that the semigroup $\rme _+^{-\rmi Ht/\hbar }$ is 
actually a retarded propagator. Similarly, the semigroup 
$\rme _-^{-\rmi Ht/\hbar }$ is actually an advanced propagator.

The following proposition, whose proof can be found in appendix~\ref{sec:A3},
asserts the soundness of the semigroups:

\vskip0.5cm

\newtheorem*{Prop4}{Proposition~4}
\begin{Prop4} \label{Prop4} The retarded propagator $\rme _+^{-\rmi Ht/\hbar}$
is well defined and coincides with $\rme ^{-\rmi Ht/\hbar}$ when $t>0$. When 
$t<0$, $\rme _+^{-\rmi Ht/\hbar}$ is not defined.  

The advanced propagator $\rme _-^{-\rmi Ht/\hbar}$ is well defined and 
coincides with $\rme ^{-\rmi Ht/\hbar}$ when $t<0$. When $t>0$, 
$\rme _-^{-\rmi Ht/\hbar}$ is not defined.  
\end{Prop4}

\vskip0.5cm

The proof of Proposition~4 makes it clear that the semigroups 
$\rme _{\pm}^{-\rmi Ht/\hbar }$ are the result of imposing upon the group 
$\rme ^{-\rmi Ht/\hbar }$ a time asymmetric boundary condition
through an analytic continuation.

Our last objective in this section is to obtain the time evolution of the
analytically continued bras and kets. Admittedly, we shall fall short of this
last objective, because at present time we only have formal results. 

By definition~(\ref{beigenrhsp}), the time evolution of the bras should
formally read as
\begin{eqnarray}
       \langle ^{\pm}q|\rme ^{-\rmi Ht/\hbar}|\varphi ^{\pm}\rangle &=&
      \langle  ^{\pm}q|  \rme ^{\rmi Ht/\hbar}\varphi ^{\pm} \rangle
             \nonumber \\
     &=& \widehat{\varphi}^{\pm}(q;-t)  \nonumber \\
     &=&   \rme ^{\rmi q^2 \hbar t/(2m)} \widehat{\varphi}^{\pm}(q) 
          \nonumber \\
     &=&
   \rme ^{\rmi q^2 \hbar t/(2m)} \langle ^{\pm}q|\varphi ^{\pm}\rangle  \, . 
         \label{deftebra}
\end{eqnarray} 
By definition~(\ref{keigenrhsc}), the time evolution of the kets
should formally read as
\begin{eqnarray}
       \langle \varphi ^{\pm}|\rme ^{-\rmi Ht/\hbar}|q^{\pm}\rangle &=&
      \langle \rme ^{\rmi Ht/\hbar}\varphi ^{\pm}|q^{\pm}\rangle
                   \nonumber \\
     &=& \overline{\widehat{\varphi}^{\pm}(\overline{q};-t)} \nonumber \\
    &=& \overline{\rme ^{\rmi \overline{q}^2\hbar t/(2m)}
        \widehat{\varphi}^{\pm}(\overline{q})} \nonumber \\
    &=& \rme ^{-\rmi q^2\hbar t/(2m)} 
        \overline{\widehat{\varphi}^{\pm}(\overline{q})} \nonumber \\
    &=&
   \rme ^{-\rmi q^2 \hbar t/(2m)} \langle \varphi ^{\pm}|q^{\pm}\rangle  \, .
         \label{defteket}
\end{eqnarray}
Plugging the limits~(\ref{timelimit0}) and (\ref{timelimitinfty}) into
Eqs.~(\ref{deftebra}) and (\ref{defteket}) should yield
\begin{equation}
      \rme ^{-\rmi Ht/\hbar}|q^{\pm}\rangle =  \rme ^{-\rmi q^2 \hbar t/(2m)}
        |q^{\pm}\rangle  \, ,
     \begin{array}{c}
             \quad t>0 \, , \  q\in {\rm 2nd, 4th} \, , \\
                         {\rm or} \\ 
              \quad t<0 \, , \ q\in {\rm 1st, 3rd} \, ,
     \end{array}
     \label{timevoketspm1} 
\end{equation}
and
\begin{equation}
      \langle ^{\pm}q|\rme ^{-\rmi Ht/\hbar} =  \rme ^{\rmi q^2 \hbar t/(2m)}
       \langle ^{\pm}q| \, , 
           \begin{array}{c}
               \quad t<0 \, , \ q\in {\rm 2nd, 4th} \, ,  \\ 
                                {\rm or} \\
               \quad t>0 \, , \ q\in {\rm 1st, 3th} \, . 
           \end{array}
      \label{timevobraspm2}
\end{equation}
The rigorous proof of Eqs.~(\ref{timevoketspm1}) and (\ref{timevobraspm2}) 
through Eqs.~(\ref{deftebra}) and (\ref{defteket}) is still 
lacking, because the invariance properties of ${\mathbf \Phi}_{\rm exp}$ under 
$\rme ^{-\rmi Ht/\hbar}$ are still not known. Such rigorous proof should 
involve a generalization of the Paley-Wiener Theorem XII~\cite{PW}, and of 
logarithmic-integral techniques~\cite{HILLE,KOOSIS}. 

One may wonder what happens to the semigroup time evolution when we make a 
complex wave number $q$ tend to a real wave number $k$. Let us do so, e.g., 
for $q$ in the fourth quadrant:
\begin{equation}
      \lim_{q\to k} \rme ^{-\rmi Ht/\hbar}|q^{\pm}\rangle = 
      \lim_{q\to k} \rme ^{-\rmi q^2 \hbar t/(2m)} |q^{\pm}\rangle =
       \rme ^{-\rmi k^2 \hbar t/(2m)} |k^{\pm}\rangle \, ,  \quad t >0 \, .
\end{equation}
It is clear from this equation that
the time evolution of $|q^{\pm}\rangle$, which should be defined for
$t>0$ only, tends to the time evolution of $|k^{\pm}\rangle$ for $t>0$. Of
course, for $t<0$, the time evolution of $|k^{\pm}\rangle$ is also defined,
even though one cannot obtain it from the above limit, since for negative
times the time evolution of $|q^{\pm}\rangle$ should not be defined.

\subsection{The ``free'' propagators}

The ``free'' time evolution $\rme ^{-\rmi H_0t/\hbar}$ can be analytically 
continued in much the same manner as $\rme ^{-\rmi Ht/\hbar}$, and such
continuation also produces semigroups. The continuation of 
$\rme ^{-\rmi H_0t/\hbar}$ into the fourth quadrant yields the
following ``free'' retarded propagator:
\begin{equation}
    \hskip-1cm   \varphi (r;t)= 
        \left( \rme _+^{-\rmi H_0t/\hbar}\varphi \right) (r)=
       \int_{\gamma _\varepsilon} \rmd q\, \rme ^{-\rmi q^2 \hbar t/(2m)}
       \widehat{\varphi}(q) \chi _0(r;q)  \, , \quad t>0 \, ,
       \label{timevownwfAC+se0} 
\end{equation}
whereas the continuation into the third quadrant yields the
following ``free'' advanced propagator:
\begin{equation}
     \hskip-1.5cm   \varphi (r;t)= 
      \left( \rme _-^{-\rmi H_0t/\hbar}\varphi \right) (r)= -
       \int_{\gamma _\varepsilon ^*} \rmd q\, \rme ^{-\rmi q^2 \hbar t/(2m)}
       \overline{\widehat{\varphi}(\overline{q})} 
        \  \overline{\chi _0(r;\overline{q})}  \, , \quad t<0 \, .
       \label{timevownwfAC-se0} 
\end{equation}
The proof that the semigroups~(\ref{timevownwfAC+se0}) and 
(\ref{timevownwfAC-se0}) are well defined follows the same steps as the proof 
of Proposition~4.

As well, the time evolution of the ``free'' bras and kets should read as
\begin{equation}
      \rme ^{-\rmi H_0t/\hbar}|q\rangle =  \rme ^{-\rmi q^2 \hbar t/(2m)}
        |q\rangle  \, ,
     \begin{array}{c}
             \quad t>0 \, , \  q\in {\rm  2nd, 4th} \, , \\ 
                            {\rm or}   \\
              \quad t<0 \, , \ q\in {\rm  1st, 3rd} \, ,
     \end{array}
     \label{timevoketspm10} 
\end{equation}
and
\begin{equation}
      \langle q|\rme ^{-\rmi H_0t/\hbar} =  \rme ^{\rmi q^2 \hbar t/(2m)}
       \langle q| \, , 
           \begin{array}{c}
               \quad t<0 \, , \ q\in {\rm 2nd, 4th} \, ,  \\ 
                                   {\rm or} \\
               \quad t>0 \, , \ q\in {\rm 1st, 3th} \, . 
           \end{array}
      \label{timevobraspm20}
\end{equation}

\section{The $\pm \rmi \varepsilon$ and time asymmetry}
\setcounter{equation}{0}
\label{sec:pmvarepos}

The Lippmann-Schwinger equation
\begin{equation}
    |E^{\pm}\rangle =|E\rangle + \frac{1}{E-H\pm \rmi \varepsilon}V |E\rangle
\end{equation}
incorporates the infinitesimal imaginary parts $\pm \rmi \varepsilon$. In 
practical calculations, $\varepsilon$ is assumed to be small, and it is made 
zero at the end of the calculation. Mathematically, the $\pm \rmi \varepsilon$
correspond to approaching the physical spectrum (the ``cut'') either from 
above ($+$) or from below ($-$). 

It has been suggested~\cite{BKW} that the $\pm \rmi \varepsilon$ should 
appear in the time evolution of the Lippmann-Schwinger kets, 
\begin{equation}
   \rme ^{-\rmi Ht/\hbar}|E^{\pm}\rangle =
     \rme ^{-\rmi (E\pm \rmi \varepsilon) t/\hbar}|E^{\pm}\rangle  \, , 
     \label{faetime}
\end{equation}
which would result in a time asymmetric evolution for the Lippmann-Schwinger 
kets. Due to $\varepsilon \neq 0$ in~(\ref{faetime}), the time evolution of 
$|E^{+}\rangle$ would be defined for $t<0$ only, and the time evolution of 
$|E^{-}\rangle$ would be defined for $t>0$ only. Thus, the time evolution of 
the Lippmann-Schwinger bras and kets associated with real energies would be 
already time asymmetric, even though no analytic continuation has been done.

However, the semigroups~(\ref{faetime}) are in conflict with the results of 
Ref.~\cite{LS1} and with standard scattering theory~\cite{TAYLOR,NUSSENZVEIG}, 
where the time evolution of the Lippmann-Schwinger bras and kets is valid for 
$-\infty < t < \infty$.

To solve this conflict, we write the Lippmann-Schwinger equation as
\begin{equation}
       |E^{\pm}\rangle = |E^{\pm}\rangle _{\rm inc} + 
                       |E^{\pm}\rangle_{\rm scattering} \, , 
\end{equation}
where
\begin{equation}
       |E^{\pm}\rangle _{\rm inc} \equiv |E\rangle  
          \label{frels}
\end{equation}
represents the incident beam and
\begin{equation}
       |E^{\pm}\rangle _{\rm scattering} \equiv  
       \frac{1}{E-H\pm \rmi \varepsilon}V|E\rangle 
          \label{splS} 
\end{equation}
represents the scattered beam. Clearly, even if we insisted on keeping 
$\varepsilon$ finite to obtain
a semigroup time evolution, the incident beam~(\ref{frels}) would still have a 
group time evolution, because $\varepsilon \neq 0$ affects only the
scattered beam~(\ref{splS}). Therefore, the semigroups~(\ref{faetime}) 
are not associated with the Lippmann-Schwinger equation for real energies.

\section{Conclusions}
\setcounter{equation}{0}
\label{sec:conclusions}

We have obtained and characterized the analytic continuation of the
Lippmann-Schwinger bras and kets. We have seen that the analytically continued
Lippmann-Schwinger bras and kets are distributions that act on the space of 
test functions ${\mathbf \Phi}_{\rm exp}$. The elements of 
${\mathbf \Phi}_{\rm exp}$ fall off at infinity like $\rme ^{-r^2}$, and in 
the wave number representation they grow like $\rme ^{|{\rm Im}(q)|^2}$. 

We have also constructed the wave number representation of the analytically
continued bras and kets, $\langle ^{\pm}\widehat{q}|$ and 
$|\widehat{q}^{\pm}\rangle$. When their associated eigenfunction does 
not have a pole, $\langle ^{\pm}\widehat{q}|$ and 
$|\widehat{q}^{\pm}\rangle$ act, respectively, as the linear and antilinear 
complex delta functional. When their associated eigenfunction has a pole, 
$\langle ^{\pm}\widehat{q}|$ and $|\widehat{q}^{\pm}\rangle$ act,
respectively, as the linear and antilinear residue functional. There is, in 
particular, a 1:1 correspondence between bras and kets for any complex wave 
number $q$.

We have proved that the analytic continuation of the time evolution of the
wave functions entails the 
imposition of a time asymmetric boundary condition. The resulting time 
evolution is given by a semigroup, which physically is simply a (retarded or 
advanced) propagator. These semigroup propagators appear as the result of 
boundary conditions, rather than as the result of an external bath. Also,
we have argued, although not fully proved, that the time evolution of the
analytically continued Lippmann-Schwinger bras and kets is given by 
semigroups.

These results have important consequences in resonance theory, as will
be shown elsewhere.

\ack

It is a great pleasure to acknowledge many fruitful conversations with
Alfonso Mondrag\'on over the past several years. Additional discussions
with J.~G.~Muga, M.~Gadella, A.~Bohm, I.~Egusquiza, R.~de la Llave, L.~Vega and
R.~Escobedo are also acknowledged. It is also a pleasure to acknowledge 
correspondence with Mario Rocca, who made the author aware of the
spaces of $M$ and $\Omega$ type.

This research was supported by MEC fellowship No.~SD2004-0003.


{

\appendix

\def\thesection{\Alph{section}}
\section{Useful formulas}
\setcounter{equation}{0}
\label{sec:apusfour}

Let us denote $\kappa$ by $Q$ when $\kappa$ becomes complex:
\begin{equation}
       Q\equiv Q(q) = \sqrt{\frac{2m}{\hbar ^2}(z-V_0)\, } =
        \sqrt{q^2 -\frac{2m}{\hbar ^2}V_0\, }  \, .
\end{equation}
It is then easy to check that
\begin{equation}
      \overline{Q(-\overline{q})} = - Q(q) \, ,
\end{equation}
\begin{equation}
      \overline{\sin (-\overline{q})} = -\sin (q) \, , \quad 
      \overline{\cos (-\overline{q})} = \cos (q) \, ,
\end{equation}
\begin{equation}
      \overline{{\cal J}_1(-\overline{q})} = -{\cal J}_1(q) \, , \quad 
      \overline{{\cal J}_2(-\overline{q})} = -{\cal J}_2(q)  \, ,
\end{equation}
\begin{equation}
      \overline{{\cal J}_3(-\overline{q})} = -{\cal J}_3(q) \, , \quad 
      \overline{{\cal J}_4(-\overline{q})} = -{\cal J}_4(q) \, ,
\end{equation}
\begin{equation}
      \overline{{\cal J}_{\pm}(-\overline{q})} = {\cal J}_{\pm}(q) \, ,
\end{equation}
\begin{equation}
      \overline{\chi (r;-\overline{q})} = -\chi (r;q) \, ,
\end{equation}
\begin{equation}
      \overline{\chi ^{\pm}(r;-\overline{q})} = -\chi ^{\pm}(r;q) \, . 
\end{equation}
It is also easy to check that
\begin{equation}
      Q(-q)=-Q(q) \, ,
\end{equation}
\begin{equation}
      {\cal J}_1(-q)=-{\cal J}_2(q) \, , \quad 
      {\cal J}_3(-q)=-{\cal J}_4(q) \, , 
\end{equation}
\begin{equation}
      {\cal J}_+(-q)={\cal J}_-(q) \, ,
         \label{jsejmqmps}
\end{equation}
\begin{equation}
      \chi (r;-q)=-\chi (r;q) \, ,
\end{equation}
\begin{equation}
      \chi ^+(r;-q)=-\chi ^-(r;q) \, .
\end{equation}
It is as well easy to check that
\begin{equation}
      \overline{Q(\overline{q})}=Q(q)  \, ,
\end{equation}
\begin{equation}
      \overline{\sin (\overline{q})}=\sin (q) \, , \quad 
      \overline{\cos (\overline{q})}=\cos (q) \, ,
\end{equation}
\begin{equation}
      \overline{{\cal J}_1(\overline{q})}={\cal J}_2(q) \, , \quad 
      \overline{{\cal J}_3(\overline{q})}={\cal J}_4(q) \, ,
\end{equation}
\begin{equation}
      \overline{{\cal J}_{+}(\overline{q})}={\cal J}_{-}(q) \, ,
\end{equation}
\begin{equation}
      \overline{\chi (r;\overline{q})}=\chi (r;q) \, ,
\end{equation}
\begin{equation}
      \overline{\chi ^{+}(r;\overline{q})} = \chi ^{-}(r;q) \, . 
\end{equation}
Using the above relations, one can show that
\begin{eqnarray}
       \langle r|q^{\pm}\rangle =\chi ^{\pm}(r;q) \, ,  \\
       \langle ^{\pm}q|r \rangle =\chi ^{\mp}(r;q) =
               \overline{\chi ^{\pm}(r;\overline{q})} =
               (-1)\chi ^{\pm}(r;-q) \, ,  \\
       \langle r|q\rangle =\chi _0(r;q) \, ,  \\
       \langle q|r \rangle =\chi _0(r;q) =
          \overline{\chi _0(r;\overline{q})} =
               (-1)\chi _0(r;-q) \, , \\
        \langle ^{\pm}q|r \rangle = 
           \overline{\langle r|\overline{q}^{\pm}\rangle} =
          (-1) \langle r|-q^{\pm}\rangle   \, , \\
      \langle q|r \rangle = 
           \overline{\langle r|\overline{q}\rangle} =
             (-1) \langle r|-q\rangle  \, . \\
\end{eqnarray}

\def\thesection{\Alph{section}}
\section{List of auxiliary propositions}
\setcounter{equation}{0}
\label{sec:A3}

Here we list the proofs of the propositions we stated in the paper. In
the proofs, whenever an operator $A$ is acting on the bras, we shall use
the notation $A^{\prime}$, and whenever it is acting on the kets, we shall
use the notation $A^{\times}$:
\begin{equation}
    \langle ^{\pm}q|A^{\prime}|\varphi ^{\pm} \rangle := 
    \langle ^{\pm}q|A^{\dagger} \varphi ^{\pm} \rangle  \, , 
         \qquad  \forall  \varphi ^{\pm} \in {\mathbf \Phi}_{\rm exp} \, ,
         \label{beigenrhsp}
\end{equation}
\begin{equation}
    \langle \varphi ^{\pm}|A^{\times}|q^{\pm}\rangle :=
    \langle A^{\dagger}\varphi ^{\pm}|q^{\pm}\rangle \, , 
           \qquad   \forall \varphi ^{\pm}\in {\mathbf \Phi}_{\rm exp} \, . 
       \label{keigenrhsc}
\end{equation}
Thus, $A^{\prime}$ denotes the \emph{dual} extension of $A$ acting to the 
left on the elements of ${\mathbf \Phi}_{\rm exp}^{\prime}$, whereas
$A^{\times}$ denotes the \emph{antidual} extension of $A$ acting to 
the right on the elements of ${\mathbf \Phi}_{\rm exp}^{\times}$. This 
notation stresses that $A$ is acting outside the Hilbert space and specifies 
toward what direction the operator is acting, thereby making the proofs more 
transparent.

\vskip1cm

\begin{proof}[{\bf Proof of Proposition~1}] \quad 

Equation~(\ref{jsejmqmps}) implies that any estimate
satisfied by ${\cal J}_+(q)$ in the upper (lower) half plane is automatically
satisfied by ${\cal J}_-(q)$ in the lower (upper) half plane. Thus, we
only need to prove Eqs.~(\ref{boundinjslp+}) and (\ref{boundinjsup+}).

From, for example, Eq.~(12.8) 
in Ref.~\cite{TAYLOR}, it follows that
\begin{equation}
    \left| {\cal J}_+(q)-1 \right| \leq  
    \frac{C}{|q|}\int_0^{\infty}\rmd r \, \left| V(r)\right|
    \frac{\left| qr\right|}{1+\left| qr\right|} \,
  \rme ^{[\, |{\rm Im}(q)| - {\rm Im}(q) \, ]r}  \, .
    \label{tayinjofu}
\end{equation}
Because ${\rm Im}(q)\geq 0$ when $q$ belongs to the
upper half plane $\mathbb{C}^+$, because our potential vanishes
when $r\notin (a,b)$, and 
because $\left| qr\right| < 1+\left| qr\right|$, Eq.~(\ref{tayinjofu}) leads to
\begin{eqnarray}
     \left| {\cal J}_+(q)-1 \right| &\leq &
    \frac{C}{|q|}\int_a^{b}\rmd r \, V_0
        \frac{\left| qr\right|}{1+\left| qr\right|} 
       \nonumber \\
    &<& \frac{C}{|q|}V_0 \int_a^{b}\rmd r\, 
     \nonumber \\
       &=&\frac{C}{|q|} V_0(b-a) \, ,
       \quad  q \in \mathbb{C}^+ \, ;
\end{eqnarray}
that is,
\begin{equation}
     \left| {\cal J}_+(q)-1 \right| <\frac{C}{|q|} \, ,
      \quad  q \in \mathbb{C}^+  \, .
\end{equation}
This inequality implies that the Jost function ${\cal J}_+(q)$ tends 
uniformly to $1$ as the wave number tends to infinity in the upper half 
plane. This uniform convergence means that for any $\varepsilon >0$, there 
exists an $R_{\varepsilon}>0$ such that for all $q \in \mathbb{C}^+$ 
satisfying $|q|\geq R_{\varepsilon}$, 
$\left|{\cal J}_+(q)-1\right|<\varepsilon$. Choose 
$\varepsilon =1/4$. Then, there exists an $R_4>0$ so that for all 
$q \in \mathbb{C}^+ $ satisfying $|q|\geq R_4$, ${\cal J}_+(q)$ lies 
within the disk of radius $1/4$ centered at $1$. This implies, in 
particular, that $\left|{\cal J}_+(q)\right|>1/2$ when $|q|\geq R_4$. Hence,
\begin{equation}
      \frac{1}{\left|{\cal J}_+(q)\right|}<2 \, , 
       \quad q \in \mathbb{C}^+ \, , \   
       |q|>R_4 \, .
      \label{bound1}
\end{equation}
This inequality proves that $1/{\cal J}_+(q)$ is bounded in the 
upper half-plane except for the following closed half-disk:
\begin{equation}
      D:=\left\{ q \in \mathbb{C}^+ \, | \quad |q|\leq R_4 \right\} .
\end{equation}
Because the Jost function does not vanish in $D$ for the potential we are 
considering (there is no bound state), $1/{\cal J}_+(q)$ is 
an analytic function in $D$. By the 
{\it Maximum Modulus Principle}, this analytic function is 
bounded by some $M>0$ when $q \in D$:
\begin{equation}
     \frac{1}{\left| {\cal J}_+(q) \right|} \leq M \, , \quad
      q \in D \, .
      \label{bound2}
\end{equation}
From Eqs.~(\ref{bound1}) and (\ref{bound2}), it follows that 
\begin{equation}
      \frac{1}{\left| {\cal J}_+(q) \right|} \leq 
       \max {(M,2)} \, , \quad
       q \in \mathbb{C}^+ \, ,
          \label{boundednessj}
\end{equation}
which proves Eq.~(\ref{boundinjslp+}). Note that for potentials that bind 
bound states, inequality~(\ref{boundednessj}) holds when 
$|q|>|K_{\rm ground}|$, where $K_{\rm ground}$ is the wave number of 
the ground state. 

Finally, the asymptotic behavior~(\ref{boundinjsup+}) can be found in 
Ref.~\cite{NUSSENZVEIG}, Eq.~(5.5.13).

\renewcommand{\qedsymbol}{}
\end{proof}

\begin{proof}[{\bf Proof of Proposition~2}] \quad 

({\it i}) The proof of ({\it i}) 
is straightforward. 

({\it ii}) In order to prove ({\it ii}), we need to realize that the space 
${\mathbf \Phi}_{\rm exp}$ satisfies
\begin{equation}
     C_0^{\infty}([0,\infty )/\{0,a,b\}) \subset 
     {\mathbf \Phi}_{\rm exp} \subset L^2([0,\infty ),\rmd r) \, ,
      \label{chainofinclusion}
\end{equation}
where $C_0^{\infty}([0,\infty )/\{0,a,b\})$ is the space of infinitely 
differentiable functions with compact support in $[0,\infty )$ that vanish
along with all their derivatives at $r=0,a,b$. Because 
$C_0^{\infty}([0,\infty )/\{0,a,b\})$ is dense in 
$L^2([0,\infty ),\rmd r)$~\cite{ROBERTS}, the chain of 
inclusions~(\ref{chainofinclusion}) implies that 
${\mathbf \Phi}_{\rm exp}$ is dense in $L^2([0,\infty ),\rmd r)$.

({\it iii}) The proof of ({\it iii}) uses the following inequality:
\begin{eqnarray}
    \| H\varphi ^{\pm} \| _{n,n'} &=& 
          \| (H+1)\varphi ^{\pm} - \varphi ^{\pm} \| _{n,n'}
     \nonumber \\
     &\leq & \| (H+1)\varphi ^{\pm} \| _{n,n'} + \| \varphi ^{\pm} \| _{n,n'} 
         \nonumber \\
     &= &\| \varphi ^{\pm} \| _{n,n'+1} + \| \varphi ^{\pm} \| _{n,n'} \,  . 
     \label{inestability}
\end{eqnarray}
This inequality implies that $H$ is 
$\mathbf \Phi _{\rm exp}$-continuous. There remains to prove that 
$\mathbf \Phi _{\rm exp}$ is 
stable under the action of $H$. In order to prove so, we need to prove that 
$H\varphi ^{\pm}$ belong to ${\cal D}$ and that the norms 
$\| H\varphi ^{\pm} \| _{n,n'}$
are finite for $n,n'=0,1,\ldots \, $. That $H\varphi ^{\pm}$ belong to 
${\cal D}$ is 
trivial from the definition of $\cal D$. That the norms 
$\| H\varphi ^{\pm} \| _{n,n'}$ are finite follows from 
inequality~(\ref{inestability}). This completes the proof of ({\it iii}).

({\it iv}) The kets $|q^{\pm}\rangle$ are well defined due to the 
properties satisfied by $\varphi ^{\pm}$. The kets $|q^{\pm}\rangle$
are antilinear functionals over the 
space $\mathbf \Phi _{\rm exp}$ by their own definition, 
Eq.~(\ref{LSdefinitionket+-q}). In order to prove that the kets 
$|q^{\pm}\rangle$ are continuous, we need the following inequality:
\begin{eqnarray}
     \left|\langle \varphi ^{\pm}|q^{\pm}\rangle \right|
     &\leq & \int_0^{\infty}\rmd r \, \left|\varphi ^{\pm}(r)
      \chi ^{\pm}(r;q)\right| \nonumber \\
     &\leq & \frac{C}{|{\cal J}_{\pm}(q)|} 
            \int_0^{\infty}\rmd r \, 
       \left|\varphi ^{\pm}(r) \frac{|q|r}{1+|q|r}
                       \rme^{|{\rm Im}(q)|r} \right| ,
      \label{interinque}
\end{eqnarray}
where we have used Eq.~(\ref{estimateofphi}) in the second step. If we take
the smallest positive integer $n$ such that $|q|\leq n$, then we 
have
\begin{eqnarray}
      \left| \frac{|q|r}{1+|q|r}\rme ^{|q|r} \right| &\leq &      
        \frac{nr}{1+nr}\rme ^{nr} \nonumber \\ 
       &=& \frac{nr}{1+nr}\rme ^{(n+1)r} \rme ^{-r} \nonumber \\ 
       &\leq & \frac{(n+1)r}{1+(n+1)r}\rme ^{(n+1)r} \rme ^{-r} \nonumber \\
       &\leq & \frac{(n+1)r}{1+(n+1)r}\rme ^{(n+1)r^2/2} \rme ^{-r+2n+2} \, .  
      \label{incresingchofwefu}
\end{eqnarray}
Plugging this inequality into (\ref{interinque}) yields
\begin{eqnarray}
   \hskip-2cm   \left|\langle \varphi ^{\pm}|q^{\pm} \rangle \right|
     &\leq & \frac{C}{|{\cal J}_{\pm}(q)|} \int_0^{\infty}\rmd r \, 
       \left|\varphi ^{\pm}(r) 
      \frac{(n+1)r}{1+(n+1)r}\rme ^{(n+1)r^2/2} \right| \, \rme ^{-r+2n+2} 
     \nonumber \\
     &\leq & \frac{C \, \rme ^{2n+2} }{|{\cal J}_{\pm}(q)|} 
             \left( \int_0^{\infty}\rmd r \, 
       \left|\varphi ^{\pm}(r) 
      \frac{(n+1)r}{1+(n+1)r}\rme ^{(n+1)r^2/2} \right| ^2 \right)^{1/2} 
      \left( \int_0^{\infty}\rmd r \, \rme ^{-2r} \right)^{1/2} \nonumber \\
    &=& \frac{C \, \rme ^{2n+2} }{|{\cal J}_{\pm}(q)|} 
                 \ \| \varphi ^{\pm} \|_{n+1,0}  \, .
      \label{interinqueb}
\end{eqnarray}
This inequality proves that the functionals $|q^{\pm} \rangle$ are
$\mathbf \Phi _{\rm exp}$-continuous except when
$q \in Z_{\pm}$. When $q \in Z_{\pm}$, one can obtain the same result by
substituting $\chi ^{\pm}(r;q)$ by their residues at $q$.

We note in passing that the same arguments lead to the following 
inequality:
\begin{equation}
      \left| \left( 1+\frac{\hbar ^2}{2m}q^2\right)^{n'}
       \langle \varphi ^{\pm}|q^{\pm}\rangle \right| \leq 
       \frac{C \, \rme ^{2n+2} }{|{\cal J}_{\pm}(q)|}   
                 \,  \| \varphi ^{\pm} \| _{n+1,n'}
     \, , \quad n'=0,1, \ldots \, . 
\end{equation}

({\it v}) We prove ({\it v}) by integration by parts and by 
using the Gaussian falloff of the functions $\varphi ^{\pm}(r)$ at infinity 
and the fact that they vanish at the origin:
\begin{eqnarray}
       \langle \varphi ^{\pm}|H^{\times}|q^{\pm}\rangle &=& 
      \langle H\varphi ^{\pm}|q^{\pm}\rangle \nonumber  \\
      &=& \int_0^{\infty}\rmd r \, 
       \left( -\frac{\hbar ^2}{2m}\frac{\rmd ^2}{\rmd r^2}+V(r) \right)
       \overline{\varphi ^{\pm}(r)} \chi ^{\pm}(r;q) \nonumber \\
       &=&- \frac{\hbar ^2}{2m}
       \left[ \frac{\rmd \overline{\varphi ^{\pm}(r)}}{\rmd r} 
          \chi ^{\pm}(r;q) \right] _0^{\infty} 
       +\frac{\hbar ^2}{2m}
      \left[ \overline{\varphi ^{\pm}(r)} \frac{\rmd \chi ^{\pm}(r;q)}{\rmd r} 
       \right] _0^{\infty} \nonumber \\ 
       &&+ \int_0^{\infty}\rmd r \, \overline{\varphi ^{\pm}(r)}
       \left( -\frac{\hbar ^2}{2m}\frac{\rmd ^2}{\rmd r^2}+V(r) \right)
        \chi ^{\pm}(r;q)
        \nonumber \\
       &=& \int_0^{\infty}\rmd r \, \overline{\varphi ^{\pm}(r)}
       \left( -\frac{\hbar ^2}{2m}\frac{\rmd ^2}{\rmd r^2}+V(r) \right)
        \chi ^{\pm}(r;q) \nonumber \\
         &=& \frac{\hbar ^2}{2m} q^2 \int_0^{\infty}
          \rmd r \, \overline{\varphi ^{\pm}(r)}
           \chi ^{\pm}(r;q)  \nonumber  \\
       &=& \frac{\hbar ^2}{2m} q^2 \, \langle \varphi ^{\pm}|q^{\pm}\rangle 
                    \, .
       \label{equaioniekets}
\end{eqnarray}

({\it vi}) That the bras are continuous can be shown through the
following inequality:
\begin{equation}
  \left|\langle ^{\pm}q|\varphi ^{\pm} \rangle \right|
     \leq    \frac{C \, \rme ^{2n+2}}{|{\cal J}_{\mp}(q)|}  
              \ \| \varphi ^{\pm} \|_{n+1,0}  \, ,
      \label{interinquebbra}
\end{equation}
where $n$ is the smallest positive integer such that $|q|\leq n$. The proof 
of~(\ref{interinquebbra}) is almost identical to the proof 
of~(\ref{interinqueb}).

({\it vii}) Equation~(\ref{kpssleftkeofHb}) can be proved in an almost 
identical manner to Eq.~(\ref{keigeeqbis}).
\renewcommand{\qedsymbol}{}
\end{proof}

\begin{proof}[{\bf Proof of Proposition~3}] \quad

The proofs of Eqs.~(\ref{boundinwhvalp+})-(\ref{boundinwhvaup-}) all
follow the same pattern, and hence we shall only need to prove 
Eq.~(\ref{boundinwhvalp+}).

When ${\rm Im}(q) \leq 0$, we have that
\begin{eqnarray}
     \hskip-2.2cm  \left| \left( 1+\frac{\hbar ^2}{2m}q^2\right)^{n'} 
         \widehat{\varphi} ^+(q)\right| &=  
     \left| \int_0^{\infty}\rmd r \, \chi ^-(r;q) (1+H)^{n'}
            \varphi ^+(r)\right|
     & \mbox{by (\ref{wnrofHamil})}  \nonumber \\
          &\leq  \int_0^{\infty}\rmd r \, 
                 \left|\chi ^-(r;q) (1+H)^{n'}\varphi ^+(r)\right| 
          &    \quad   \nonumber \\
          &\leq  C  \int_0^{\infty}\rmd r \, 
                 \left| \rme ^{|{\rm Im}(q)|r} (1+H)^{n'}
                     \varphi ^+(r) \right|   
              &  \mbox{by (\ref{boundrschi-lower})}  \nonumber \\
          &\leq  C  \int_0^{\infty}\rmd r \, 
      \left|\rme ^{|{\rm Im}(q)|^2/(2\alpha )} \rme ^{\alpha r^2/2} 
       (1+H)^{n'} \varphi ^+(r)  \right|   
              &  \mbox{by (\ref{pqss2n})}     \nonumber \\
          & = C \, \rme ^{|{\rm Im}(q)|^2/(2\alpha )}
        \int_0^{\infty}\rmd r \, 
       \left| \rme ^{-\alpha r^2/2} \rme ^{\alpha r^2} 
                (1+H)^{n'} \varphi ^+(r) \right|
           &  \hskip1cm \quad      \nonumber \\
         &\leq  C \,  \rme ^{|{\rm Im}(q)|^2/(2\alpha )}
 \left( \int_0^{\infty}\rmd r \, 
         \left| \rme ^{-\alpha r^2/2}  \right|^2 \right) ^{1/2} 
        &    \quad      \nonumber \\ 
       & \quad  \times        
     \left( \int_0^{\infty}\rmd r \, 
       \left|\rme ^{\alpha r^2} (1+H)^{n'}
                    \varphi ^+(r) \right|^2 \right) ^{1/2} 
           &   \quad   \nonumber \\ 
       &= C \, \rme ^{|{\rm Im}(q)|^2/(2\alpha )} 
          \left( \int_0^{\infty}\rmd r \, 
     \left| \rme ^{\alpha r^2} (1+H)^{n'} 
                     \varphi ^+(r)\right|^2 \right) ^{1/2} . 
         & \quad 
         \label{firstineq}
\end{eqnarray}
There only remains to prove that the last integral is finite. In order to
prove so, we split that integral into two:
\begin{eqnarray}
 \int_0^{\infty}\rmd r \, 
         \left|\rme ^{\alpha r^2} (1+H)^{n'}\varphi ^+(r) \right|^2  &=
   \int_0^{1}\rmd r \, 
         \left|\rme ^{\alpha r^2} (1+H)^{n'}\varphi ^+(r) \right|^2 
      \nonumber \\  
   & \quad  +  \int_1^{\infty}\rmd r \, 
         \left| \rme ^{\alpha r^2} (1+H)^{n'}\varphi ^+(r) \right|^2 
     \nonumber \\  
   & \equiv I_1 + I_2 \, .
\end{eqnarray}
Now, on the one hand,
\begin{eqnarray}
     I_1 & = \int_0^{1}\rmd r \, 
         \left|\rme ^{\alpha r^2} (1+H)^{n'}\varphi ^+(r) \right|^2 
      \nonumber \\  
   & \leq \rme ^{\alpha}  \int_0^{1}\rmd r \, 
         \left| (1+H)^{n'}\varphi ^+(r) \right|^2 
      \nonumber \\  
   & \leq \rme ^{\alpha}  \int_0^{\infty}\rmd r \, 
         \left| (1+H)^{n'}\varphi ^+(r) \right|^2 
      \nonumber \\  
   & = \rme ^{\alpha} \,  \| (1+H)^{n'}  \varphi ^+ \|^2  \, ,
\end{eqnarray}
which is finite, since $\varphi ^+$ belongs, in particular, to the
maximal invariant subspace of $H$, see Eq.~(\ref{condition1}). On the other 
hand, if we take $n$ as the smallest positive integer that is larger than
$2$ and $2\alpha$, then
\begin{eqnarray}
     I_2 &= \int_1^{\infty}\rmd r \, 
         \left|\rme ^{\alpha r^2} (1+H)^{n'}\varphi ^+(r) \right|^2 
     \nonumber \\  
   & \leq \frac{9}{4} \int_1^{\infty}\rmd r \, 
         \left| \frac{2r}{1+2r}\rme ^{\alpha r^2}(1+H)^{n'}
                       \varphi ^+(r)\right|^2
      \nonumber \\ 
   & \leq \frac{9}{4} \int_1^{\infty}\rmd r \, 
         \left| \frac{nr}{1+nr}\rme ^{nr^2/2}(1+H)^{n'}\varphi ^+(r) \right|^2
      \nonumber \\   
   & \leq \frac{9}{4} \int_0^{\infty}\rmd r \, 
         \left| \frac{nr}{1+nr}\rme ^{nr^2/2}(1+H)^{n'}\varphi ^+(r)\right|^2 
       \nonumber \\  
   & = \frac{9}{4} \,  \| \varphi ^+ \| _{n,n'}^2 \, ,
        \label{lastineq}
\end{eqnarray}
where in the last step we have used definition~(\ref{normsLS}). The
combination of Eqs.~(\ref{firstineq})-(\ref{lastineq}) yields the
estimate~(\ref{boundinwhvalp+}).

\renewcommand{\qedsymbol}{}
\end{proof}

\begin{proof}[{\bf Proof of Proposition~4}] \quad

The proof of~(\ref{timevownwfAC-se}) is very similar to the proof 
of~(\ref{timevownwfAC+se}), and therefore we shall only prove the latter.

We just need to prove that for $\varepsilon >0$ and $t>0$, it holds that
\begin{equation}
      \hskip-2cm  
     \varphi ^{\pm}(r;t)=
       \int_0^{\infty} \rmd k\, \rme ^{-\rmi k^2\hbar t/(2m)}
       \widehat{\varphi}^{\pm}(k) \chi ^{\pm} (r;k) =
     \int_{\gamma _{\varepsilon}} \rmd q\, \rme ^{-\rmi q^2\hbar t/(2m)}
       \widehat{\varphi}^{\pm}(q) \chi ^{\pm} (r;q) \, .
         \label{defomrfoqua}
\end{equation}

Equation~(\ref{defomrfoqua}) can be easily proved after proving that the 
integrand on the right hand side tends to zero in the limit $|q|\to \infty$ 
while the argument of $q$ remains within $0$ and $\varepsilon$. In 
order to prove so, we write the complex wave number as 
$q= |q|\rme ^{-\rmi \theta}$, $0\leq  \theta  \leq \varepsilon$, and use the 
estimates of Propositions~1 and~3 for large $q$:
\begin{eqnarray}
      \hskip-2cm
       \left|\rme ^{-\rmi q^2\hbar t/(2m)}
       \widehat{\varphi}^{\pm}(q) \chi ^{\pm}(r;q)\right| & = & 
        \rme ^{-|q|^2 \sin (2\theta)\hbar t/(2m)}
       \left| \widehat{\varphi}^{\pm}(q) \chi ^{\pm}(r;q) \right| \nonumber \\
       &\leq& C
        \rme ^{-|q|^2 \sin (2\theta)\hbar t/(2m)}
        \rme ^{\frac{|q|^2 \sin ^2\theta}{2\alpha}}
        \left| \chi ^{\pm}(r;q) \right| \nonumber \\
        &\leq& C 
         \rme ^{-|q|^2 \sin (2\theta)\hbar t/(2m)}
        \rme ^{\frac{|q|^2 \sin ^2\theta}{2\alpha}}
           \frac{|q|r}{1+|q|r}  \rme ^{|q|r \sin \theta}   \, .
         \label{beatfnofint} 
\end{eqnarray}
As $|q|$ tends to infinity, the exponential that carries the time 
dependence dominates if we choose 
$\alpha > m/(2\hbar t) \tan \varepsilon$. Thus,
when $t>0$ and $0\leq \theta \leq \varepsilon$, Eq.~(\ref{beatfnofint}) tends 
to zero uniformly when the argument of $q$ belongs to $[0,\varepsilon ]$: 
\begin{equation}
      \left| \rme ^{-\rmi q^2\hbar t/(2m)}
       \widehat{\varphi}^{\pm} (q) \chi ^{\pm} (r;q)\right|  
        \mapupdown{{\rm uniformly}}{\quad |q|\to \infty \quad} 0 \, , \quad 
       \theta \in [0,\varepsilon ]  \, .
     \label{integtezerinf}
\end{equation}

With help from this limit, it is very easy to prove Eq.~(\ref{defomrfoqua}). We
first consider the contour $\Gamma _R$, which consists of the segment
$[0, R]$, the arc $\gamma _R$ of radius $R$ that sweeps in between the angles 
$0$ and $\varepsilon$, and the segment $\gamma _{\varepsilon, R}$ of length
$R$ that links the origin with the lower end of $\gamma _R$, see 
Fig.~\ref{fig:gammas}b. Then, by Cauchy's theorem, we have that
\begin{equation}
       \int_{\Gamma _R} \rmd q\, \rme ^{-\rmi q^2\hbar t/(2m)}
       \widehat{\varphi}^{\pm}(q) \chi ^{\pm}(r;q) = 0 \, ,
         \label{Gamma1}
\end{equation}
because the integrand is analytic inside 
$\Gamma _R$. Disassembling~(\ref{Gamma1}) yields
\begin{eqnarray}
     \hskip-1cm
       \int_{\gamma _{\varepsilon ,R}} \rmd q\, \rme ^{-\rmi q^2\hbar t/(2m)}
       \widehat{\varphi}^{\pm}(q) \chi ^{\pm}(r;q) &-& 
       \int_0^{R} \rmd k\, \rme ^{-\rmi k^2\hbar t/(2m)}
       \widehat{\varphi}^{\pm}(k) \chi ^{\pm}(r;k)  \nonumber \\ 
       &-& 
       \int_{\gamma _{R}} \rmd q\, \rme ^{-\rmi q^2\hbar t/(2m)}
       \widehat{\varphi}^{\pm}(q) \chi ^{\pm}(r;q) =0  \, .
           \label{Gamma2}
\end{eqnarray}
Because of~(\ref{integtezerinf}), the third integral in
Eq.~(\ref{Gamma2}) vanishes as $R$ tends to infinity. Thus, taking the 
limit $R\to \infty$ of Eq.~(\ref{Gamma2}) yields the sought 
result~(\ref{defomrfoqua}).
\renewcommand{\qedsymbol}{}
\end{proof}

\def\thesection{\Alph{section}}
\section{$M$ and $\Omega$ functions}
\setcounter{equation}{0}
\label{sec:omegafunc}

In this appendix, we collect some results on $M$ and $\Omega$ functions
from Chapter~I of Ref.~\cite{GELFANDIII}.

Let $\mu (\xi)$ ($0\leq \xi < \infty$) denote an increasing continuous 
function, such that $\mu (0)=0$, $\mu (\infty )=\infty$. We define for
$x\geq 0$
\begin{equation}
       M(x)=\int_0^{x}\rmd \xi \, \mu (\xi) \, .
       \label{defiM}
\end{equation}
The function $M(x)$ is an increasing convex continuous function, with
$M(0)=0$, $M(\infty )=\infty$.

Let $\omega (\eta )$ ($0\leq \eta < \infty$) denote an increasing continuous 
function, with $\omega (0)=0$, $\omega (\infty )=\infty$. For $y\geq 0$ we 
define
\begin{equation}
       \Omega (y)=\int_0^{y}\rmd \eta \, \omega (\eta ) \, .
       \label{defiOme}
\end{equation}
The function $\Omega (y)$ is an increasing convex continuous function, with
$\Omega (0)=0$, $\Omega (\infty )=\infty$.

We now introduce the important concept of functions which
are {\it dual in the sense of Young}. Let the functions $M(x)$ and 
$\Omega (y)$ be defined by Eqs.~(\ref{defiM}) and (\ref{defiOme}), 
respectively. If the functions $\mu (\xi)$ and $\omega (\eta )$ which occur
in these equations are mutually inverse, i.e., 
$\mu[\omega (\eta )]=\eta$, $\omega [\mu (\xi )]=\xi$, then the corresponding
functions are said to be {\it dual in the sense of Young}. In this case,
the Young inequality
\begin{equation}
     xy\leq M(x) +\Omega (y) 
        \label{dissentangle}
\end{equation}
holds for any $x , y \geq 0$, see Ref.~\cite{GELFANDIII}. The Young 
inequality ``disentangles'' the product $xy$ into the sum of a function
that depends only on $x$ and a function that depends only on $y$. 

As an application of Eq.~(\ref{dissentangle}), one can prove that
\begin{equation}
     xy\leq \frac{x^a}{a} +\frac{y^b}{b} \, , 
        \label{pqss}
\end{equation}
where $a$ and $b$ are real numbers satisfying
\begin{equation}
            \frac{1}{a} +\frac{1}{b} = 1 \, . 
        \label{pqand}
\end{equation}
When $a=b=2$, we get
\begin{equation}
     xy\leq \frac{x^2}{2} +\frac{y^2}{2} \, , 
        \label{pqss2}
\end{equation}
which yields the following inequality for any $\alpha >0$:
\begin{equation}
     xy\leq \alpha \frac{x^2}{2} + \frac{1}{\alpha}\frac{y^2}{2}  \, . 
        \label{pqss2n}
\end{equation}

}


\newpage

\begin{figure}[ht]
\hskip-0.5cm \includegraphics{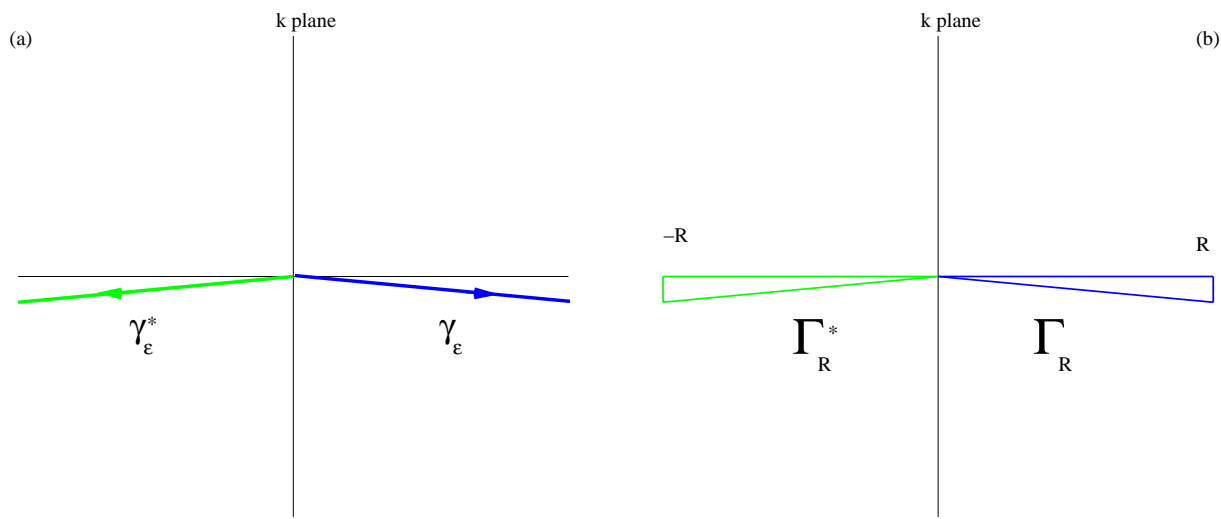}
\caption{The contour $\gamma _{\varepsilon}$ is a radial path in the fourth
quadrant that forms an angle $-\varepsilon$ with the positive $k$-axis. The 
contour $\Gamma _R$ consists of the segment $[0,R]$ of the 
positive real line, the arc $\gamma _{R}$, and the segment
$\gamma _{\varepsilon, R}$ of length $R$ that forms an angle $-\varepsilon$ 
with the positive $k$-axis. The contours $\gamma _{\varepsilon}^*$ and 
$\Gamma _R^*$ 
are the mirror images of $\gamma _{\varepsilon}$ and $\Gamma _R$ with respect 
to the imaginary axis. If necessary, $\gamma _{\varepsilon}$ and 
$\gamma _{\varepsilon}^*$ may be bent to avoid resonances.}
\label{fig:gammas}
\end{figure}

\end{document}